\documentclass[12pt]{iopart}
\usepackage{epsfig,amssymb,psfig,graphicx,bm}

\def\spose#1{\hbox to 0pt{#1\hss}}
\def\lesssim{\mathrel{\spose{\lower 3pt\hbox{$\mathchar"218$}}
 \raise 2.0pt\hbox{$\mathchar"13C$}}}
\def\gtrsim{\mathrel{\spose{\lower 3pt\hbox{$\mathchar"218$}}
 \raise 2.0pt\hbox{$\mathchar"13E$}}}

\def\<{\langle}
\def\>{\rangle}
\def\phm{\hphantom{$-$}}

\begin{document}

\title{Magnetic and glassy transitions in the square-lattice XY model with
   random phase shifts}

\author{Vincenzo Alba$^1$, Andrea Pelissetto$^2$ and Ettore Vicari$^3$}
\address{$^1$ Scuola Normale Superiore and INFN, I-56126 Pisa, Italy}
\address{$^2$ Dipartimento di Fisica dell'Universit\`a di Roma ``La Sapienza"
        and INFN, I-00185 Roma, Italy}
\address{$^3$ Dipartimento di Fisica dell'Universit\`a di Pisa
        and INFN, I-56127 Pisa, Italy}

\ead{
Vincenzo.Alba@sns.it,
Andrea.Pelissetto@roma1.infn.it,
Ettore.Vicari@df.unipi.it}

\begin{abstract}

  We investigate the magnetic and glassy transitions of the square-lattice XY
  model in the presence of random phase shifts.  We consider two different
  random-shift distributions: 
  the Gaussian distribution
  and a slightly different distribution (cosine distribution) 
  which allows the exact determination of the Nishimori line where
  magnetic and overlap correlation functions are equal.
  We perform Monte Carlo simulations for several values of the temperature and 
  of the variance of the disorder distribution,
  in the paramagnetic phase close to the magnetic
  and glassy transition lines. We find that, along the transition line 
  separating the paramagnetic and the quasi-long-range order phases,
  magnetic correlation functions show a universal Kosterlitz-Thouless behavior
  as in the pure XY model, while overlap correlations show a 
  disorder-dependent critical behavior. This behavior is observed 
  up to a multicritical point which, in the cosine model, lies on the Nishimori
  line. Finally, for large values of the disorder variance, we observe
  a universal zero-temperature glassy critical transition, which is in
  the same universality class as that occurring in the gauge-glass model.

\end{abstract}

%%%%% \pacs{PACS Numbers: 74.81.Fa, 05.70.Jk, 64.60.Fr}
%74.81.Fa Josephson junction arrays
%64.60.-i general studies of phase transitions
%64.60.Fr Equilibrium properties near critical points
%75.10.Hk classical spin models
%74.78.-w       Superconducting films and low-dimensional structures
%05.10.Ln       Monte Carlo methods
%05.70.Jk       Critical point phenomena 

\maketitle

% ========================= BODY =========================
%\narrowtext

\section{Introduction}
\label{intro}

The two-dimensional XY model with random phase shifts (RPXY) describes the
thermodynamic behavior of several disordered systems, such as Josephson
junction arrays with geometrical disorder~\cite{GK-86,GK-89}, magnetic systems
with random Dzyaloshinskii-Moriya interactions~\cite{RSN-83}, crystal systems
on disordered substrates~\cite{CF-94}, and vortex glasses in high-$T_c$
cuprate superconductors~\cite{FTY-91}.  See \cite{Korshunov-06,KR-03}
for recent reviews.  The RPXY model is defined by the partition function
\begin{eqnarray}
&&Z(\{A\}) = \exp (-{\cal H}/T),\nonumber\\
&&{\cal H} = -\sum_{\langle xy \rangle } {\rm Re} \,\bar\psi_x U_{xy} \psi_y
= - \sum_{\langle xy \rangle} {\rm cos}(\theta_x - \theta_y-A_{xy}),
\label{RPXY} 
\end{eqnarray}
where $\psi_x\equiv e^{i\theta_x}$, $U_{xy}\equiv e^{i A_{xy}}$, and the
sum runs over the bonds ${\langle xy \rangle }$ of a square lattice.  The
phases $A_{xy}$ are uncorrelated quenched random variables with zero
average.  In most studies they
are distributed with Gaussian probability
\begin{equation}
P_G(A_{xy}) \propto \exp\left(-{A_{xy}^2\over 2\sigma}\right).
\label{GRPXYd}
\end{equation}
We denote the RPXY model with distribution (\ref{GRPXYd})
by GRPXY.  We also consider the RPXY model with
distribution (cosine model)
\begin{equation}
P_C(A_{xy})\propto \exp\left({{\rm cos} A_{xy}\over \sigma}\right),
\label{CRPXYd}
\end{equation}
which we denote by CRPXY.  Such a model is particularly interesting because the
distribution (\ref{CRPXYd}) allows some exact calculations along the so-called
Nishimori (N) line $T \equiv 1/\beta=\sigma$~\cite{ON-93,Nishimori-02}.  In
both GRPXY and CRPXY models the pure XY model is recovered in the limit
$\sigma\rightarrow 0$, while the so-called gauge glass model~\cite{ES-85} with
uniformly distributed phase shifts is obtained in the limit $\sigma\rightarrow
\infty$.

The nature of the different phases arising when varying the temperature $T$
and the disorder parameter $\sigma$ and the critical behavior at the phase
transitions have been investigated in many theoretical and experimental
works~\cite{RSN-83,GK-86,GK-89,CF-94,FTY-91,Korshunov-06,KR-03,%%
  ON-93,Nishimori-02,ES-85,FLTL-88,CD-88,%%
  FBL-90,HS-90,RTYF-91,Li-92,Gingras-92,DWKHG-92,RY-93,%%
  Korshunov-93,NK-93,Nishimori-94,NSKL-95,CF-95,%%
  JKC-95,HWFGY-95,KN-96,Tang-96,BY-96,MG-97,%%
  Scheidl-97,KS-97,KCRS-97,MG-98,CL-98,Granato-98,
  SYMOUK-98,MW-99,KA-99,CP-99,%%
  Kim-00,CL-00,AK-02,HO-02,KY-02,Katzgraber-03,%%
  HKM-03,CTH-03,NW-04,KC-05,TT-05,UKMCL-06,%%
  YBC-06,CLL-08,APV-09}. In spite of that, a conclusive picture of the
phase diagram and of the critical behaviors has not been achieved yet.

\begin{figure}[tpb]
\centerline{\psfig{width=9truecm,angle=0,file=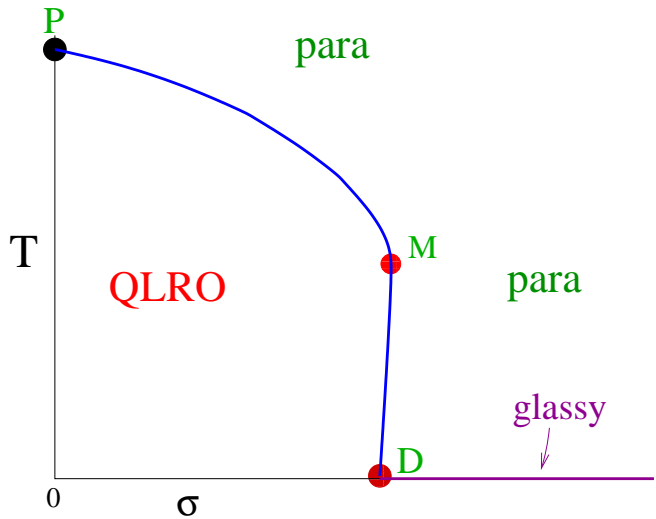}}
\caption{
Phase diagram of RPXY models as a function of 
$T$ and of the disorder-distribution variance $\sigma$. }
\label{phdia}
\end{figure}

The expected $T$-$\sigma$ phase diagram for the GRPXY and CRPXY models, which
is sketched in Fig.~\ref{phdia}, presents two finite-temperature phases: a
paramagnetic phase and a low-temperature phase characterized by 
quasi-long-range order (QLRO) for sufficiently small values of $\sigma$; see,
e.g., \cite{APV-09} and references therein.  The paramagnetic phase is
separated from the QLRO phase by a transition line, which starts from the pure
XY point (denoted by $P$ in Fig.~\ref{phdia}) at $(\sigma=0,T=T_{XY}\approx
0.893)$ and ends at a zero-temperature disorder-induced transition denoted by
$D$ at $(\sigma_D,T=0)$. The QLRO phase extends up to a maximum value
$\sigma_M$ of the disorder parameter, which is related to the point $M\equiv
(\sigma_M,T_M)$, where the tangent to the transition line is parallel to the
$T$ axis.  No long-range glassy order can exist at finite temperature for any
value of $\sigma$, including the gauge-glass limit
$\sigma\to\infty$~\cite{NK-93,Nishimori-94}.  Several numerical studies of the
gauge-glass XY
model~\cite{FTY-91,RY-93,Granato-98,AK-02,KY-02,Katzgraber-03,NW-04,KC-05,TT-05}
support a zero-temperature glassy transition. A more complete discussion of
the known features of the phase diagram will be reported below.

In this paper we investigate several controversial issues
concerning the critical behavior at the magnetic and glassy
transitions in RPXY models. In particular,
we will check whether 
the critical behavior along the paramagnetic-QLRO transition line 
is universal and belongs to the universality class
of the pure XY model, whether there is a multicritical point along the
paramagnetic-QLRO transition line,
and, finally, whether the $T=0$ glassy transition extends from
$\sigma=\infty$ to $\sigma=\sigma_D$, see Fig.~\ref{phdia}, and belongs to the
same universality class as that in the XY gauge-glass model.
For this purpose, we perform Monte
Carlo (MC) simulations of the GRPXY and CRPXY models for several values of the
temperature and of the variance $\sigma$, 
approaching the magnetic and glassy transition lines from the 
paramagnetic phase. As we shall see, our results for the CRPXY model 
provide a robust evidence 
for a universal Kosterlitz-Thouless
(KT) behavior of the magnetic correlations along the 
paramagnetic-QLRO transition line
from the pure XY point $P$ to the point $M$ where the
transition line runs parallel to the $T$ axis and magnetic and overlap 
correlations are equal. Along the line
the magnetic correlation length $\xi$ 
behaves as $\ln \xi \sim u_t^{-1/2}$, where $u_t$ is the thermal scaling
field, and the magnetic susceptibility as $\chi\sim \xi^{7/4}$ (corresponding
to $\eta=1/4$).  On the other hand, the
behavior of the overlap correlations appears to be $\sigma$ dependent along
this transition line.  Moreover, the numerical results for the 
CRPXY model indicate that the point $M$ is 
multicritical. We conjecture that these conclusions hold for any RPXY model.
In all cases we expect that the paramagnetic-QLRO transition line 
is divided into two parts by a multicritical point $M$, where 
magnetic and overlap correlations have the same critical behavior,
though they are not equal. At variance with what happens in the CRPXY 
model, the point $M$ is not expected to coincide with the point in 
which the tangent to the transition line is parallel to the $T$ axis: this 
coincidence should be a unique feature of the CRPXY model. Then, from $P$
to $M$ we expect any RPXY model to behave as the CRPXY, that is 
a KT behavior for magnetic correlations and a $\sigma$
dependent behavior for disorder-related quantities. 
The universality of the behavior has been confirmed by our numerical 
results for the GRPXY model. 

Finally, we have investigated the critical behavior for large values of 
$\sigma$.  Our numerical results provide strong
evidence for a universal zero-temperature glassy transition for
$\sigma>\sigma_D$. For $T\to 0$ overlap correlation functions are 
critical, and, in particular, the corresponding correlation length $\xi_o$
diverges as $\xi_o\sim T^{-\nu}$ when $T\to 0$ with $\nu=2.5(1)$.

This paper is organized as follows.  In Sec.~\ref{phadia} we review the known
results for the phase diagram and for the critical behavior of the RPXY models.
Sec.~\ref{notations} provides the definitions of the quantities considered in
our numerical work.  In Sec.~\ref{paraqlro} we study the critical behavior along
the thermal paramagnetic-QLRO transition line which starts at the pure XY 
point $P$ and ends at 
multicritical point $M$.  In Sec.~\ref{HTresnline} we discuss 
critical behavior along the N line of the CRPXY model and show that 
the point $M$ where the N line intersects the critical line 
is multicritical. In Sec.~\ref{glassybeh} we investigate the glassy
critical behavior at $T=0$ for $\sigma
>\sigma_D$.  Finally, in Sec.~\ref{conclusions} we draw our conclusions.
There are also several appendices. \ref{AppMC} reports some details
of the MC simulations.  \ref{AppRGsigmazero} is devoted to a careful
analysis of the KT renormalization-group (RG) equations and of the corresponding
RG flow. We derive the most general form of the $\beta$ function for the 
sine-Gordon model and discuss the structure of the scaling corrections
in the XY model. These results are used in the discussion of the behavior at the
paramagnetic-QLRO transition. In \ref{Appirrelevant} we discuss some
features of the critical behavior at a multicritical point.  In
\ref{AppRGdisordine} we briefly discuss the RG equations in the presence
of randomness.  Finally, in \ref{App-magn} we report some 
analytical results for the
magnetic correlations in the gauge-glass model.

\section{The phase diagram}
\label{phadia}

In Fig.~\ref{phdia} we show the expected $T$-$\sigma$ phase diagram of the
RPXY models.  In the absence of disorder ($\sigma=0$) the model shows a
high-$T$ paramagnetic phase and a low-$T$ phase characterized by 
QLRO controlled by a line of Gaussian fixed points, where the
spin-spin correlation function $\langle \bar{\psi}_x \psi_y \rangle$ decays as
$1/r^{\eta(T)}$ for $r\equiv |x-y|\to \infty$, with $\eta$ depending on $T$.
The two phases are separated by a Kosterlitz-Thouless (KT)
transition~\cite{KT-73} at \cite{HP-97} $\beta_{XY}\equiv
1/T_{XY}=1.1199(1)$. For $\tau \equiv
T/T_{XY}-1\rightarrow 0^+$, the correlation length and the
magnetic susceptibility diverge exponentially 
as ${\rm ln} \xi \sim \tau^{-1/2}$ and $\chi\sim
\xi^{7/4}$, respectively.  An interesting question is whether these features
change in the presence of random phase shifts.

The low-temperature phase of RPXY models shows QLRO for sufficiently small
values of $\sigma$.  The universal features of the long-distance behavior are
explained by the random spin-wave theory~\cite{RSN-83}, obtained by replacing
\begin{equation}
{\rm cos}(\theta_x - \theta_y-A_{xy})\longrightarrow
1 - {1\over 2} (\theta_x - \theta_y+A_{xy})^2
\label{swr}
\end{equation}
in Hamiltonian (\ref{RPXY}).  This scenario has been accurately verified by
Monte Carlo (MC) simulations in both GRPXY and CRPXY models~\cite{APV-09}.
The QLRO phase disappears for large values of $\sigma$,
see, e.g., \cite{Korshunov-06} and references therein; more precisely, as
we shall see, for $\sigma\gtrsim 0.31$ in the case of the CRPXY model.

For $\sigma\rightarrow \infty$ phases are uniformly distributed and 
one obtains the gauge-glass model. Even if this model has been much
investigated~\cite{FTY-91,ES-85,HS-90,RTYF-91,Li-92,Gingras-92,%%
  DWKHG-92,RY-93,NK-93,Nishimori-94,JKC-95,HWFGY-95,BY-96,KS-97,%%
  KCRS-97,MG-98,Granato-98,KA-99,CP-99,Kim-00,AK-02,HO-02,KY-02,%%
  Katzgraber-03,HKM-03,CTH-03,NW-04,KC-05,TT-05,UKMCL-06,CLL-08},
its phase diagram and critical behavior are still
controversial.  No long-range glassy order can exist at finite
temperature~\cite{NK-93,Nishimori-94}.  However, this does not exclude more
exotic low-temperature glassy phases~\cite{CP-99,HKM-03}, for example a phase
characterized by glassy QLRO.  Many numerical works at finite and zero
temperature support a zero-temperature transition
\cite{FTY-91,RY-93,Granato-98,AK-02,KY-02,Katzgraber-03,NW-04,KC-05,TT-05}.
According to this scenario, the correlation length determined from the overlap
correlation function diverges as $\xi_o\sim T^{-\nu}$ when approaching the
critical point $T=0$.  The critical exponent $\nu$ has been estimated by
finite-temperature Monte Carlo (MC) simulations, obtaining~\cite{KY-02}
$1/\nu=0.39(3)$ and ~\cite{NW-04} $1/\nu=0.36(3)$.  
The exponent $\nu$ is related to the $T=0$ stiffness exponent
$\theta$ by $\theta=-1/\nu$.  The $T=0$ numerical
calculations of \cite{AK-02} and \cite{TT-05} provided the estimates
$\theta=-0.36(1)$ and $\theta\approx -0.45$ respectively, which are consistent with
the finite-temperature estimates of $\nu$.  The $T=0$ transition scenario has
been questioned in
\cite{CP-99,Kim-00,HO-02,HKM-03,CTH-03,UKMCL-06,YBC-06,CLL-08}, which
provide some numerical and experimental (for Josephson-junction arrays with
positional disorder~\cite{YBC-06}) evidence for the existence of a
finite-temperature transition at $T\approx 0.2$, with a low-temperature glassy
phase characterized by frozen vortices and glassy QLRO.

Other features of the phase diagram are better discussed within the CRPXY
model, characterized by the random phase-shift distribution (\ref{CRPXYd}),
because of the existence of exact results along the so-called Nishimori (N)
line~\cite{ON-93,Nishimori-02}
\begin{equation}
T \equiv 1/\beta=\sigma.
\label{Nline}
\end{equation}
Along the N line the energy density $E$ is known exactly:
\begin{equation}
E\equiv {1\over V} [\langle {\cal H} \rangle] = -2{I_1(\beta)\over I_0(\beta)},
\label{exe}
\end{equation} 
where $I_0(\beta)$ and $I_1(\beta)$ are modified Bessel functions.
Moreover, the spin-spin and overlap correlation
functions are equal:
\begin{eqnarray}
[\langle \bar{\psi}_x \psi_y \rangle ] = 
[|\langle \bar{\psi}_x \psi_y \rangle|^2 ].
\label{cnl}
\end{eqnarray}
As already noted in \cite{Nishimori-02}, the N line should play an
important role in the phase diagram, because it is expected to mark the
crossover between the magnetic-dominated region 
and the disorder-dominated one. 

In the GRPXY and CRPXY models, the paramagnetic phase is separated from the
magnetic QLRO phase by a transition line, which starts from the pure XY
point (denoted by $P$ in Fig.~\ref{phdia}) at $(\sigma=0,T=T_{XY}\approx
0.893)$ and ends at a $T=0$ transition point induced by disorder (denoted by
$D$) at $(\sigma_D,T=0)$, where $\sigma_D>0$.\footnote{We mention that the
  first renormalization-group (RG) analyses based on a Coulomb-gas
  description~\cite{RSN-83} predicted $\sigma_D = 0$, but it was later
  clarified that this was an artefact of the approximations. Indeed,
  experimental~\cite{FLTL-88} and numerical
  works~\cite{FLTL-88,CD-88,FBL-90,MG-97} as well as refinings of the RG
  arguments~\cite{ON-93,NSKL-95,KN-96,Tang-96,Scheidl-97,CL-98}, have shown
  the absence of a reentrant transition for sufficiently small values of
  $\sigma$.}
An important result has been proven for the CRPXY model~\cite{ON-93}: the
critical value $\sigma_M$ of $\sigma$ along the N line is an upper bound for
the values of $\sigma$ where magnetic QLRO can exist.  Therefore, 
at the critical point $M\equiv
(\sigma_M,T_M)$ the tangent to the critical line should be parallel to the 
$T$ axis; moreover, the critical
value $\sigma_D$ at $T=0$ must satisfy $\sigma_D\le\sigma_M$.

\begin{figure}[tpb]
\begin{center}
\psfig{width=9truecm,angle=0,file=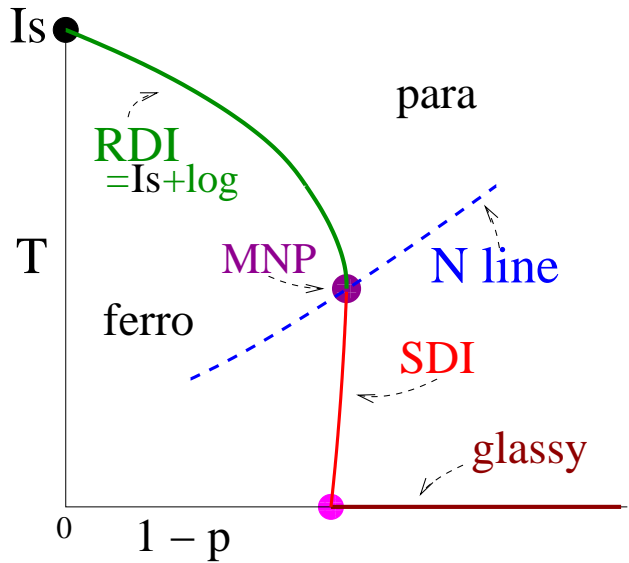}
\caption{Phase diagram of the $\pm J$ (Edwards-Anderson) Ising model on the
  square lattice.
The phase diagram is symmetric under $p\rightarrow 1-p$.}
\label{phdiaisi2d}
\end{center}
\end{figure}

It is worth noting how similar the phase diagrams of the CRPXY model and of the
square-lattice $\pm J$ Ising model in the $T$-$p$ plane are, see
Figs.~\ref{phdia} and \ref{phdiaisi2d}, respectively.  The square lattice $\pm
J$ (Edwards-Anderson) Ising model is defined by the Hamiltonian
\begin{equation}
{\cal H}_{\pm J} = - \sum_{\langle xy
    \rangle} J_{xy} \sigma_x \sigma_y,
\label{pmj}
\end{equation}
where $\sigma_x=\pm 1$, the sum is over pairs of nearest-neighbor sites of a
square lattice, and $J_{xy}$ are uncorrelated quenched random variables,
taking values $\pm J$ with probability distribution $P(J_{xy}) = p
\delta(J_{xy} - J) + (1-p) \delta(J_{xy} + J)$.  This model presents an
analogous N line~\cite{Nishimori-81} in the $T$-$p$ phase diagram, defined by
$\tanh (1/T) - 2 p + 1=0$.  The transition point along
the N line is a multicritical point (MNP)
\cite{HPPV-08,PHP-06}. Moreover, the critical behavior for $T>T_{MNP}$ and 
$T<T_{MNP}$ is different.
From the pure Ising point at $p=1$ to the MNP the
critical behavior is analogous to that observed in 2D randomly dilute Ising
(RDI) models \cite{HPPV-08-2}. From the MNP to the $T=0$ axis the
critical behavior belongs to a new strong-disorder Ising (SDI) universality
class \cite{PPV-08}. Finally, the $T=0$ end-point of the low-temperature
paramagnetic-ferromagnetic transition
line is the starting point of a $T=0$ transition line, characterized by a glassy
universal critical behavior~\cite{PPV-10}.

In \cite{ON-93} it was also argued that, in the RPXY models 
(in particular, in the CRPXY one)
the low-temperature paramagnetic-QLRO transition line from the critical point
$M$ to the point $D$ runs parallel to the $T$ axis,
so that $\sigma_D=\sigma_M$.  The same arguments fail in the 2D $\pm J$ Ising
model~\cite{PPV-08,HPPV-08,PHP-06,AH-04,WHP-03}, although they provide a good
approximation.  Thus, they are likely not exact also 
in the case of the RPXY models,
although they may still provide a good approximation, suggesting that $0 < 
\sigma_M-\sigma_D\ll \sigma_M$.

In the phase diagram reported in Fig.~\ref{phdia}, which refers to the CRPXY,
we may distinguish two
transition lines meeting at point $M$: the thermal paramagnetic-QLRO
transition line from $P$ to $M$, which can be approached by decreasing the
temperature at fixed $\sigma$, and the transition line from $M$ to $D$, which
can be instead observed by changing disorder at fixed $T$ for sufficiently low
temperatures. As we shall see, our numerical results for the 
CRPXY model provide some evidence that the point $M$ is multicritical.
We conjecture that the same conclusion holds for generic RPXY models, 
though in the generic case we do not expect the multicritical point $M$
to coincide with the point where the tangent to the critical line
is parallel to the $T$ axis.

The phase transition from the paramagnetic to the QLRO phase is generally
expected to be of KT type ($\ln \xi$ is expected to have a power-law
divergence), but its specific features, for instance the precise form 
of the power-law behavior and the value of the exponent $\eta$, have
not been conclusively determined yet. 
Some numerical results supporting the KT-like behavior
were presented in \cite{MG-97}.  The disorder-driven $T=0$ transition at
$\sigma_D$ has been argued~\cite{NSKL-95,CF-95,MG-97,CL-98,CL-00} to show a
KT-like behavior with $\ln \xi \sim (\sigma-\sigma_D)^{-1}$ and $\chi\sim
\xi^{2-\eta}$ with $\eta=1/16$.  However, other
RG studies \cite{Scheidl-97,Tang-96} obtained a different 
behavior: $\ln \xi \sim (\sigma-\sigma_D)^{-1/2}$.
The value of $\eta$ associated with the
magnetic two-point function has been believed to vary along the critical
line~\cite{RSN-83,NSKL-95,Tang-96,Scheidl-97}, from $\eta=1/4$ of the pure XY
model at $\sigma=0$ to $\eta=1/16$ at the $T=0$ transition.  As we shall see,
our numerical results along the thermal paramagnetic-QLRO transition line, 
from $P$ to and including $M$, strongly support $\eta=1/4$, independently
of $\sigma$.

In the following sections we investigate some of the open issues of the RPXY
models, by performing MC simulations of the GRPXY and CRPXY models close to 
their magnetic and glassy transition lines. In particular, we investigate the
critical behavior at the thermal paramagnetic-QLRO transition line (from point 
$P$ to the multicritical point), along the
N line in the CRPXY model, and at the $T=0$ glassy transition line for large
disorder.

\section{Notations}
\label{notations}

We consider RPXY models defined on square lattices of size $L^2$ with periodic
boundary conditions.  We define the magnetic spin-spin correlation function
\begin{equation}
G(x-y) \equiv [ \langle \bar{\psi}_x \,\psi_y \rangle ]
\label{magcorr}
\end{equation}
and the overlap correlation function
\begin{equation}
G_o(x-y) \equiv [ |\langle \bar{\psi}_x \,\psi_y \rangle|^2 ].
\label{overcorr}
\end{equation}
The angular and square brackets indicate the thermal average and the
quenched average over disorder, respectively. 
The latter can also be written in terms of the overlap variables.
Given two copies of the system with spins $\psi^{(1)}_x$ and 
$\psi^{(2)}_x$, we define 
\begin{equation}
q_x = \bar{\psi}_x^{(1)} \psi_x^{(2)},\qquad
G_o(x-y) = [ \langle \bar{q}_x \,q_y \rangle ],
\label{overlap}
\end{equation}
where the thermal average is performed over the two systems with the same
disorder configuration. We 
define the magnetic susceptibility $\chi\equiv \sum_x G(x)$, the overlap
susceptibility $\chi_o\equiv \sum_x G_o(x)$, and the second-moment correlation
lengths
\begin{eqnarray}
\xi^2 \equiv {\widetilde{G}(0) - 
\widetilde{G}(q_{\rm min}) \over 
          \hat{q}_{\rm min}^2 \widetilde{G}(q_{\rm min}) },
\qquad 
\xi_{o}^2 \equiv {\widetilde{G}_o(0) - 
\widetilde{G}_o(q_{\rm min}) \over 
          \hat{q}_{\rm min}^2 \widetilde{G}_o(q_{\rm min}) },
\label{smc}
\end{eqnarray}
where $q_{\rm min} \equiv (2\pi/L,0)$, $\hat{q} \equiv 2 \sin q/2$.

We also define the quartic couplings 
\begin{eqnarray}
&&g_4 \equiv - {3\chi_4\over 2\chi^2\xi^2}, 
\qquad \chi_4 \equiv {1\over V}
[ \langle |\mu|^4 \rangle - 2 \langle |\mu|^2 \rangle^2 ], 
\label{g4def} \\
&&g_{22} \equiv -{\chi_{22}\over \chi^2\xi^2},\qquad
\chi_{22}\equiv {1\over V} \left( [ \langle |\mu|^2 \rangle^2 ] - 
  [ \langle |\mu|^2 \rangle ]^2\right), 
\label{g22def}\\
&&g_c \equiv g_4 + 3 g_{22},\label{gcdef}
\end{eqnarray}
where $\mu \equiv \sum_x \psi_x$ and $V=L^2$.  
Note that for the pure XY model $g_{22}=0$ and $g_c=g_4$.
Finally, we define an overlap quartic coupling $g_{o}$ as
\begin{eqnarray}
&&g_{o} \equiv - {3 \bar\chi_{4o}
\over 2\chi_o^2\xi_o^2},
\qquad\bar\chi_{4o} = {1\over V} [ \langle |\mu_o|^4 \rangle] - 2
[\langle |\mu_o|^2 \rangle ]^2 ,
\label{g4odef} 
\end{eqnarray}
where $\mu_o \equiv \sum_x q_x$.

\section{Critical behavior along the thermal para-QLRO transition line}
\label{paraqlro}

\begin{figure}[tpb]
\begin{center}
\psfig{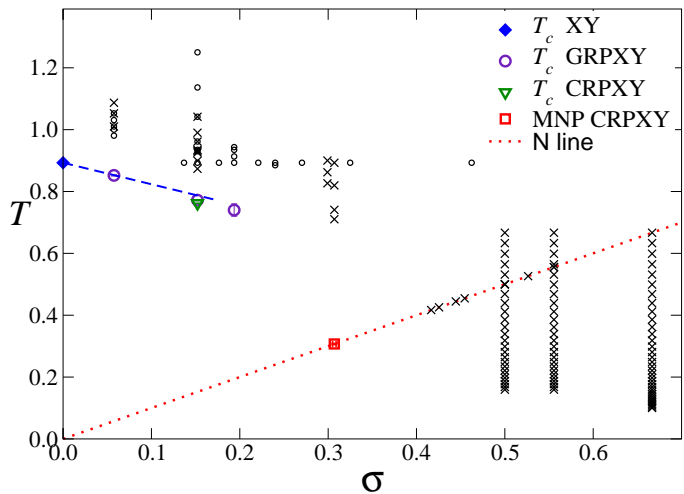}
\caption{ Values of $T\equiv 1/\beta$ and $\sigma$ where MC data were
  collected. The circles and crosses refer to the GRPXY and CRPXY models,
  respectively.  The dotted line $T = \sigma$ is the N line for the CRPXY
  model. We also show some estimates of $T_c$ for the GRPXY and CRPXY models,
  and the critical point (MNP) of the CRPXY model along the N line. The dashed
  line is the prediction (\ref{taupred}) for the behavior of $T_c$ at small
  values of $\sigma$.  }
\label{tc}
\end{center}
\end{figure}

In this section we study the critical behavior of the RPXY models along the
thermal paramagnetic-QLRO transition line, see Fig.~\ref{phdia}, which starts
at the point $P$ on the $\sigma=0$ axis and ends at the multicritical 
point, which
belongs to the N line in the CRPXY model.  
For this purpose, we perform MC simulations of the GRPXY and of the CRPXY
model for several values of $T$ and $\sigma$ in the paramagnetic phase, where
the magnetic correlation length $\xi$ is large but finite.  Fig.~\ref{tc}
shows the points where the simulations are performed.  The MC algorithm is
described in \ref{AppMC}.  We average over a large number of samples,
$N_s\approx 10^4$ in most cases.  We consider large lattice sizes,
satisfying $L/\xi\gtrsim 10$, in order to make finite-size effects negligible
and obtain infinite-volume results.  The residual finite-size
effects are in all cases smaller than, or at most comparable with, the
statistical errors.

In the following we first discuss the critical behavior of the magnetic
spin-spin correlation function (\ref{magcorr}).  We show that disorder is
apparently irrelevant: for any $\sigma$ the correlation length diverges
following the KT law valid for $\sigma = 0$ and the magnetic susceptibility
diverges with critical exponent $\eta$ equal to 1/4.  Then, we discuss the
behavior of observables related to the overlap correlation function
(\ref{overcorr}), finding that the critical behavior of these quantities is
apparently $\sigma$ dependent.

\subsection{Critical behavior approaching the pure XY transition point}
\label{XYpuro}

We wish now to understand the critical behavior along any line that lies in
the paramagnetic phase and ends at the pure XY critical point at $\sigma = 0$
and $T = T_{XY}$.  For $\sigma = 0$, as $T$ approaches the critical
temperature $T_{XY}$ from above (paramagnetic phase), the magnetic correlation
length $\xi$ diverges as
\begin{equation}
{\rm ln} (\xi/X) = C \tau^{-1/2} + O(\tau^{1/2}),
\qquad \tau\equiv (T-T_{XY})/T_{XY},
\label{KTb2}
\end{equation}
where $X$ and $C$ are nonuniversal constants.  In the case of the
square-lattice XY model with nearest-neighbor interactions~\cite{HP-97}
$\beta_{XY}\equiv 1/T_{XY}=1.1199(1)$, $X=0.233(3)$ and 
$C = 1.776(4)$.\footnote{Equation~(\ref{KTb2}) holds 
  whatever the definition of the correlation
  length is, but of course $X$ depends on the specific choice for $\xi$.
  Reference \cite{HP-97} studied the exponential correlation length $\xi_{\rm
    gap}$, which is defined as the inverse of the mass gap, and determined the
  corresponding constant $X_{\rm gap} = 0.233(3)$.  Since in the critical
  limit \cite{CPRV-96} $\xi^2/\xi_{\rm gap}^2= r=0.9985(5)$, the constant $X$
  for the second-moment correlation length we use is given by $X = X_{\rm gap}
  \sqrt{r} = 0.233(3)$.}  The magnetic susceptibility $\chi$ diverges as, see
\ref{AppRGsigmazero},
\begin{equation}
\chi = A_\chi \xi^{7/4} \left[ 1 + {b_\chi\over 
{\rm ln} (\xi/X)} +  O\left( 1/{\rm ln}^2\xi\right)\right].
\label{chibeh}
\end{equation}
Note that while $A_\chi$ is a nonuniversal amplitude, the coefficient $b_\chi$
of the leading logarithmic corrections is universal.  As shown in
\ref{AppRGsigmazero}, it can be computed from the perturbative expansion
of the RG dimension of the spin variable, obtaining $b_\chi=\pi^2/16$.

\begin{figure}[tpb]
\begin{center}
\psfig{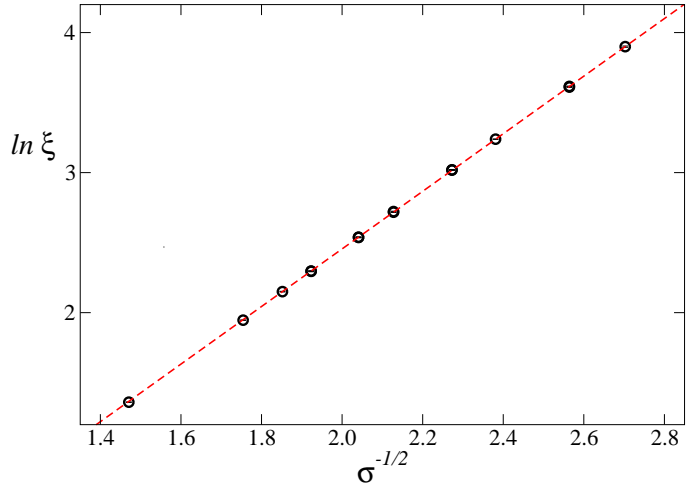}
\caption{ MC estimates of $\xi$ for $\beta=\beta_{XY}=1.1199$ 
  and several values of $\sigma$ versus $\sigma^{-1/2}$.  The dashed line
  corresponds to a linear fit to ${\rm ln} \xi = C_\sigma \sigma^{-1/2} + b$.
}
\label{xiTxy}
\end{center}
\end{figure}

We now consider the GRPXY model and study the critical behavior of $\chi$ and
$\xi$ as one approaches the pure XY critical point along the line $\beta =
\beta_{XY} = 1.1199$ by decreasing $\sigma$.  We collected data for
$0.46\gtrsim \sigma\gtrsim 0.14$ in the infinite-volume limit, corresponding
to the quite large range of correlation lengths $4\lesssim \xi\lesssim 50$.
Fig.~\ref{xiTxy} shows a plot of $\ln \xi$ versus $\sigma^{-1/2}$.  The data
fall on a straight line, showing that for $\sigma\to 0$
\begin{equation}
\ln \xi \sim  \sigma^{-1/2}.
\label{xivss}
\end{equation}
This behavior can be understood within the RG framework.  The general
discussion presented in \ref{Appirrelevant} shows that, as long as
disorder is less relevant than the thermal perturbation, the critical behavior
can be simply obtained by replacing $\tau$ with the nonlinear thermal scaling
field.  Note that it is not necessary that disorder is irrelevant to obtain
the result (\ref{xivss}).  In general, the thermal nonlinear scaling field $u_t$
is an analytic function of the system parameters. Thus, in the presence of
disorder it is a function of both $\tau = (T - T_{XY})/T_{XY}$ and
$\sigma$ such that, close to the XY transition point, it behaves as
\begin{equation}
  u_t(\tau,\sigma) = \tau + c_\sigma \sigma + \ldots
\label{utau}
\end{equation}
where the dots stand for higher-order terms. If disorder is less relevant 
than the thermal perturbation, then
\begin{equation}
{\rm ln} (\xi/X) = C u_t^{-1/2} + O(u_t^{1/2}),
\label{loxiut}
\end{equation}
along any straight line in the $T,\sigma$ plane which ends at the XY pure
transition point. Since this relation also holds for $\sigma = 0$ and
$u_t(\tau,0) = \tau$, $C$ and $X$ are the same constants reported below
(\ref{KTb2}). Along the line $T = T_{XY}$ Equation~(\ref{loxiut}) implies
\begin{equation}
{\rm ln} (\xi/X) = {C\over (c_\sigma \sigma)^{1/2} } + O(\sigma^{1/2}),
\label{alongkt}
\end{equation}
in agreement with the observed behavior.  In order to determine $c_\sigma$ we
have performed fits to
\begin{equation}
{\rm ln} (\xi/X) = C_\sigma \sigma^{-1/2}\left( 1 + b \sigma\right),
\label{fits}
\end{equation} 
using $X=0.233(3)$. We obtain the estimates $C_\sigma=2.010(2)$ and $b\approx
-0.11$. In particular, a fit of the data satisfying $\xi\gtrsim 7$ gives
$C_\sigma=2.0102(8)$ and $b=-0.108(2)$, with $\chi^2/{\rm DOF}\approx 1.1$
(DOF is the number of degrees of freedom of the fit).  Using $C=1.776(4)$ and
$C_\sigma=C/\sqrt{c_\sigma}$, we obtain
\begin{equation}
c_\sigma = \left({C\over C_\sigma}\right)^2 = 0.781(4).
\end{equation}  
The constant $c_\sigma$ is nonuniversal and as such is model dependent.
However, for $\sigma\to 0$ the fields $A_{xy}$ are typically very
small and the distribution functions for the GRPXY and CRPXY models are
identical to leading order in $A_{xy}$.  We thus expect that the first
correction to the thermal scaling field due to disorder is identical in the
two models, i.e.
\begin{eqnarray}
u_{t,{\rm GRPXY}}(\tau,\sigma) = u_{t,{\rm CRPXY}}(\tau,\sigma) + O(\sigma^2),
\label{utequality}
\end{eqnarray}
which implies that $c_\sigma$ is the same in the GRPXY and CRPXY models.

\subsection{Critical behavior of the magnetic correlations at fixed $\sigma$}
\label{ktmagncorr}

\begin{figure}[tpb]
\begin{center}
\psfig{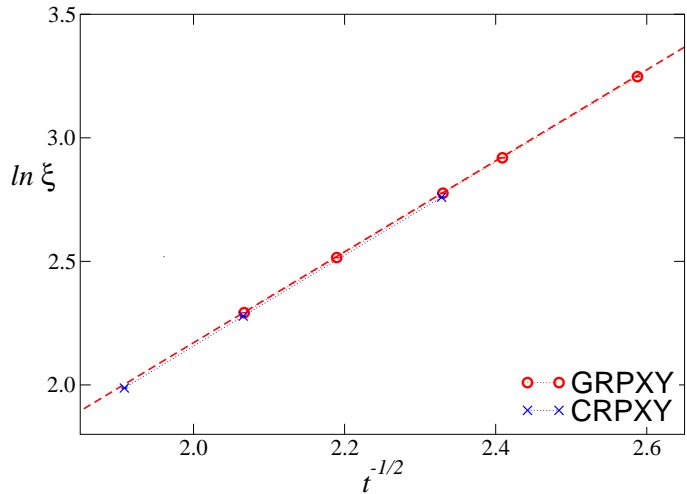}
\caption{Plots of 
  $\ln \xi$ vs $t^{-1/2}$, where $t\equiv (T-T_c)/T_c$, for the GRPXY
  and CRPXY models at $\sigma=0.0576$. For both models we use $T_c=0.8528$, as
  obtained by using (\ref{taupred}).
  The dashed line corresponds to a fit of
  the GRPXY data to ${\ln \xi} = c t^{-1/2} + a$. The
  dotted line that 
  connects the MC data is drawn to guide the eye. }
\label{xi0p24}
\end{center}
\end{figure}

Standard arguments that apply to critical lines and multicritical points
imply that the critical temperature at fixed $\sigma$ must be the solution of
the equation
\begin{equation}
u_t[T_c(\sigma),\sigma]=0. 
\label{tcequt}
\end{equation}
Therefore, Equation (\ref{utau}) also
implies that for small values of $\sigma$ the critical temperature for the
GRPXY model (and also for the CRPXY model if (\ref{utequality}) 
holds) is given by
\begin{equation}
T_c(\sigma) = T_{XY}[1 - c_\sigma \sigma + O(\sigma^2)].
\label{taupred}
\end{equation}
Equation~(\ref{taupred}) can be checked by analyzing data at fixed 
small values of
$\sigma$. We have performed MC simulations of the GRPXY model at $\sigma =
0.0576$ for several values of $\beta$, from $\beta=0.95$ to $\beta=1.02$,
corresponding to $10\lesssim \xi \lesssim 26$, and of the CRPXY model at the
same value of $\sigma$ for $\beta=0.92,\,0.95,\,0.99$ corresponding to
$7\lesssim \xi \lesssim 16$.  In Fig.~\ref{xi0p24} we plot $\xi$ versus
$t^{-1/2}$ with $t\equiv T/T_c-1$ and $T_c=0.8528$ given by
(\ref{taupred}) [if we take the errors on $T_{XY}$ and $c_\sigma$ into
account, we have $T_c=0.8528(3)$].  Clearly, $\xi\to\infty$ as $t \to 0$,
confirming (\ref{taupred}). Moreover, they are clearly consistent with the
KT behavior
\begin{equation}
{\ln \xi} = a t^{-1/2} + b.
\label{genlogfit}
\end{equation}
A fit of all available data for the GRPXY model to (\ref{genlogfit}) gives
$a=1.841(2)$ and $b=-1.511(5)$ (with $\chi^2/{\rm DOF}\approx 1.3$) keeping
$T_c=0.8528$ fixed.  A nonlinear fit, taking $T_c$ as a free parameter, gives
$T_c=0.852(2)$, in good agreement with (\ref{taupred}).  Note that the
estimate of the constant $b$ is close to the corresponding XY-model value
$\ln X=-1.46(1)$.  This is no unexpected since $X(\sigma)=X + O(\sigma)$.

\begin{figure}[tpb]
\begin{center}
\psfig{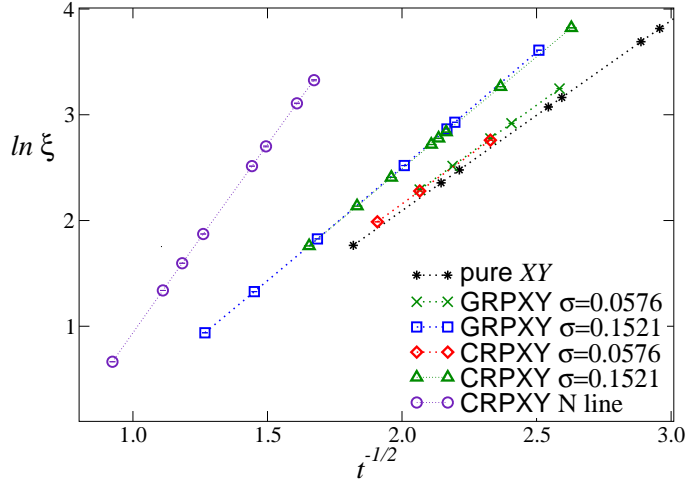}
\caption{ Estimates of $\ln \xi$ vs $t^{-1/2}$, where 
  $t\equiv T/T_c(\sigma)-1$, for
  the GRPXY and CRPXY models for several values of $\sigma$. For 
  $\sigma = 0.0576$ we take $T_c(\sigma) = 0.8528$ [Equation~(\ref{taupred})]. 
  For the other values of $\sigma$, $T_c(\sigma)$ is determined from the 
  data. The lines are drawn to guide the eye. The data for the XY are taken 
  from \cite{BNNPSW-01}.}
\label{xi}
\end{center}
\end{figure}

We also collected data at $\sigma=0.1521$ for both the GRPXY and CRPXY models,
for $0.8\le \beta\le 1.1199$ (corresponding to $2\lesssim \xi\lesssim 37$) and
$0.96\le \beta\le 1.145$ (corresponding to $5 \lesssim \xi\lesssim 46$),
respectively.  Again, the data fit well the KT behavior (\ref{genlogfit}), see
Fig.~\ref{xi}.  Fits of the MC data for $\xi\gtrsim 10$ to
(\ref{genlogfit}) (for which $\chi^2/{\rm DOF}<1$) give the estimates
$T_c=0.772(2)$ for the GRPXY model, and $T_c=0.762(1)$ for the CRPXY model.
Note that (\ref{taupred}) would give $T_c=0.7872$ for $\sigma=0.1521$,
which is slightly larger than the above estimates. This is not unexpected
since, when increasing $\sigma$, higher-order terms (which are different for
the two models) may become important in (\ref{utau}).  We also mention the
estimates $b=-1.82(7)$ and $b=-1.78(3)$ for the GRPXY and CRPXY model,
respectively, from which one obtains estimates of the corresponding length
scale $X(\sigma) = e^b$, $X = 0.162(11)$ and $X = 0.169(5)$.  We also
determined $\xi$ for other values of $\sigma$, but in a smaller range.  The
results are compatible with a KT behavior, but they do not allow us to get
robust estimates of $T_c$. We only mention that in the case of the GRPXY at
$\sigma=0.1936$, for which we have only data for $\xi\lesssim 20$, we find
$T_c \approx 0.74$.

\begin{figure}[tpb]
\begin{center}
\psfig{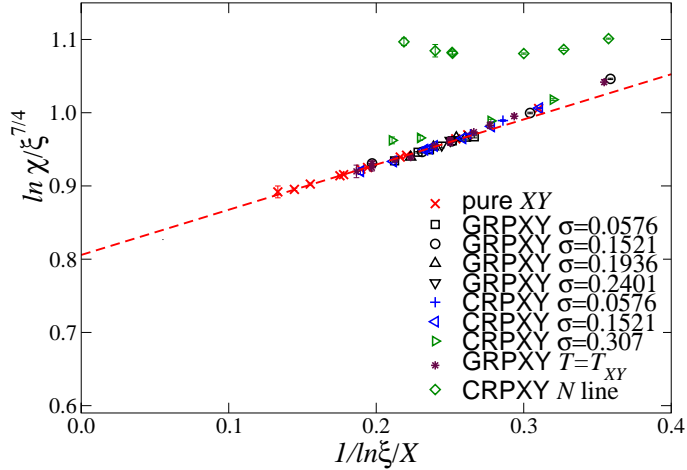}
\caption{ Plot of $\ln(\chi/\xi^{7/4})$ versus $1/\ln\xi/X$. We fix 
  $X=0.233$, which is the length-scale value  valid for the pure XY model.
We show data for the pure XY model (taken from
  \cite{BNNPSW-01}), 
  and for the GRPXY and CRPXY models at various values of
  $\sigma$, at $T=T_{XY}$ and along the N line. The dashed line 
  corresponds to a fit
  to $a+\pi^2/(16\ln\xi/X)$ of the pure-XY data satisfying
  $\xi\gtrsim 10$ (we obtain $a=0.8058(1)$ with $\chi^2/{\rm DOF}\approx 0.7$).  }
\label{eta}
\end{center}
\end{figure}

At a KT transition the magnetic susceptibility behaves as in
(\ref{chibeh}), where $b_\chi=\pi^2/16$ is universal.  In Fig.~\ref{eta}
we show $\chi/\xi^{7/4}$ for the GRPXY and CRPXY and several values of
$\sigma$ together with those of the pure XY model taken from
\cite{BNNPSW-01}. We report the data versus $\ln \xi/X(\sigma=0)$.  We
could have also used $\ln \xi/X(\sigma)$, where $X(\sigma)$ is determined from
the fit of $\xi$. This choice gives a plot essentially identical to the one
reported, which is not unexpected since, by using $\ln \xi/X(\sigma=0)$ or
$\ln \xi/X(\sigma)$ one simply changes the corrections of order $\sigma/\ln^2
\xi/X$, which are present anyway.  The results appear to follow the same curve
within the errors (except those obtained along the N line, which we shall
discuss in Sec.~\ref{HTresnline}).  They provide strong evidence that the
value $\eta=1/4$ is universal along the thermal paramagnetic-QLRO transition
line. Also the slope appears universal (the coefficient $b_\chi$ does not
depend on $\sigma$), as expected on the basis of the discussion of
\ref{AppRGsigmazero}.  The constant $A_\chi$ corresponds to the intercept
of $\chi/\xi^{7/4}$ at $\ln \xi/X(\sigma)=0$. As it can be seen from the
figure, this constant, which is not universal, varies very little with
$\sigma$: differences are not visible within our errors, except for the CRPXY 
data
at $\sigma=0.307$. However, note that for this value of $\sigma$ the critical
behavior is controlled by the multicritical Nishimori point, i.e. by the
special point $M$ which appears in Fig.~\ref{phdia}; we will return to it in
Sec.~\ref{HTresnline}.

In conclusion, the above numerical results provide a strong evidence that the
magnetic two-point correlations show a KT behavior along the thermal 
paramagnetic-QLRO transition line in GRPXY and CRPXY models.

\subsection{Quartic couplings}

We now discuss the behavior of the quartic couplings defined in 
(\ref{g4def})-(\ref{gcdef}).
We recall that in the pure XY model $g_{22}=0$ while $g_4=g_c$ behaves as
\begin{equation}
g_4 = g_4^* + {b_g\over (\ln \xi/X)^{2}} +  O(1/\ln^4 \xi),
\label{g4beh}
\end{equation}
where $g_4^*$ and $b_g$ are universal; see \ref{AppRGsigmazero}.  We
mention the estimates $g_4^*=13.65(6)$ obtained by form-factor computations in
\cite{BNNPSW-01}, and $g_4^*=13.7(2)$ by field-theoretical
methods~\cite{PV-00}; other results for $g_4^*$ can be found in \cite{PV-r}
and references therein.

\begin{figure}[tpb]
\begin{center}
\psfig{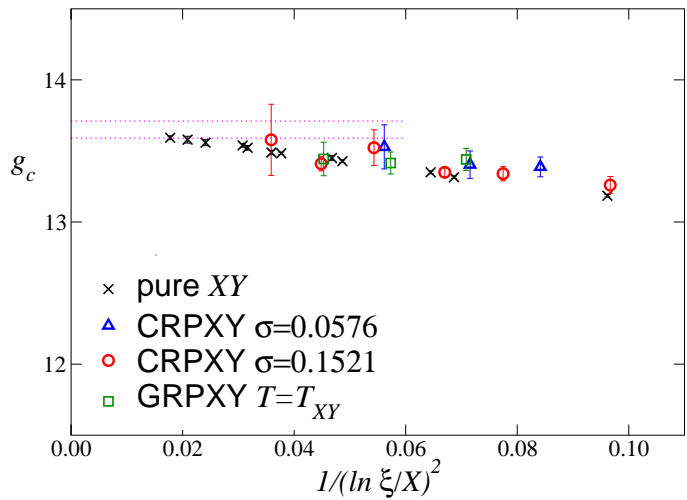}
\caption{ 
  MC estimates of $g_c\equiv g_4+3 g_{22}$ vs $1/(\ln \xi/X)^2$ with $X=0.233$.
  The data for the pure XY model are taken from
  \cite{BNNPSW-01}.  The dotted lines correspond to the estimate
  $g_c^*=g_4^*=13.65(6)$ obtained by form-factor
  calculations~\cite{BNNPSW-01}.  }
\label{g4}
\end{center}
\end{figure}

In Fig.~\ref{g4} we show some MC results of $g_c$ for the CRPXY model at
$\sigma=0.1521,\,0.0576$ and the GRPXY model at $\beta=\beta_{XY}=1.1199$
(within our errors of a few per mille the infinite-volume limit is
reached for $L/\xi\gtrsim 10$, as in the pure XY model~\cite{BNNPSW-01}),
and compare them with MC results for the pure XY model taken from
\cite{BNNPSW-01}.  The results are identical within 
errors.  For example, if we consider the CRPXY model for $\sigma=0.1521$,
a fit to $g_c^* + b_g/(\ln \xi/X)^2$ gives $g_c^*=13.57(10)$ and 
$b_g=-3.1(1.4)$, with
$\chi^2/{\rm DOF}\approx 0.4$, to be compared with the value~\cite{BNNPSW-01}
$g_4^*=13.65(6)$ of the pure XY model. Both $g_c^*$ and $b_g$,
which are universal in the pure-XY universality class, do not 
depend on $\sigma$.

\begin{figure}[tpb]
\begin{center}
\psfig{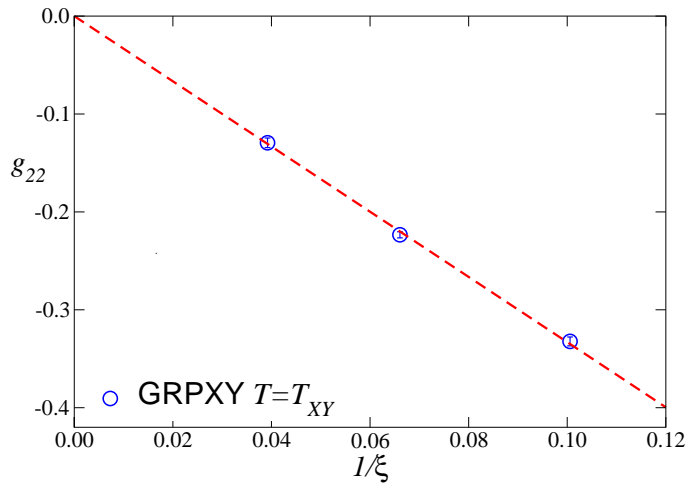}
\caption{Estimates of $g_{22}$ versus $1/\xi$ for the GRPXY model at
 fixed $\beta=\beta_{XY}=1.1199$. The line is a fit of 
 $g_{22}$ to $c \xi^{-1}$.  }
\label{g22Txy}
\end{center}
\end{figure}

The quartic coupling $g_{22}$ defined in (\ref{g22def}) is interesting
because it is particularly sensitive to randomness effects, since in the pure
XY model it vanishes trivially.  The estimates of $g_{22}$ in the GRPXY model 
for $T=T_{XY}$ and several values of $\sigma$ 
are shown in Fig.~\ref{g22Txy}. They decrease with
decreasing $\sigma$, and appear to vanish when $\sigma\to 0$ as
\begin{equation}
g_{22}\sim c \xi^{-\varepsilon},
\label{g22txy}
\end{equation}
with $\epsilon\approx 1.0$. A fit to (\ref{g22txy}) gives $\varepsilon =
0.97(4)$, $c=3.1(3)$ with $\chi^2/{\rm DOF}\approx 1.1$, where DOF is the
number of degrees of freedom of the fit. 

\begin{figure}[tpb]
\begin{center}
\psfig{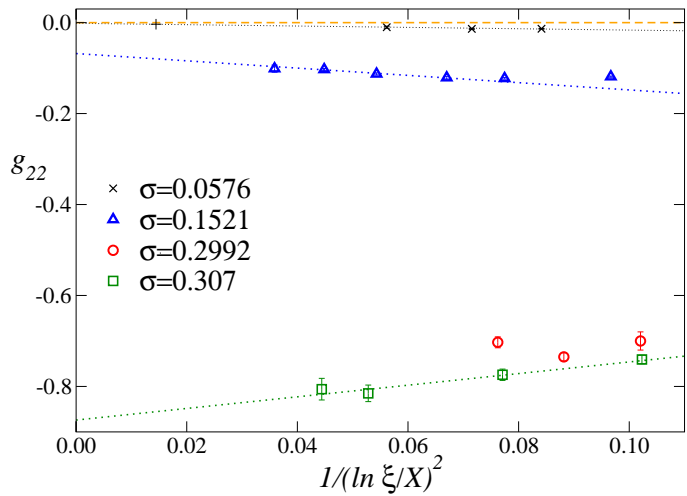}
\caption{ Estimates of $g_{22}$ for the CRPXY model
  at various values of $\sigma$. The lines show linear extrapolations to the
  critical point.  The data denoted by a plus along the line related to
  $\sigma=0.0576$ is obtained by using (\ref{g22txy}) and (\ref{fits})
  with the results of the fits along the $T=T_{XY}$ line.  }
\label{g22all}
\end{center}
\end{figure}

The fast decrease of $g_{22}$ along
the line $T=T_{XY}$ [note that $g_{22}\sim 1/\xi$ implies
$g_{22} \sim \exp(-c\sigma^{-1/2})$]
might suggest the irrelevance of
disorder, and therefore that the critical value $g_{22}^*$ vanishes along the
thermal paramagnetic-QLRO transition line.
This conclusion is apparently contradicted by the results at fixed $\sigma>0$. 
The results for the CRPXY model at various values of $\sigma$,
$\sigma=0.0576,\,0.1521,\,0.2992,\,0.307$, are shown in Fig.~\ref{g22all},
where they are plotted versus $(\ln \xi/X)^{-2}$, which is the correction
expected in the pure XY model for RG invariant quantities.  
The coupling $g_{22}$
is quite small, but definitely different from zero on the transition line.  
For $\sigma=0.1521$ an
extrapolation using $g_{22}^* + b/(\ln \xi/X)^2$ suggests a nonzero
critical limit. Using only data satisfying $\xi\gtrsim 10$, this fit gives
$g_{22}^*=-0.068(8)$ and $b=-0.080(15)$, with $\chi^2/{\rm DOF}\approx 0.4$.
We should also mention that the data for the largest values of $\xi$, those 
satisfying
$\xi\gtrsim 10$ say, may be consistent with a vanishing critical limit, but
only assuming a slower logarithmic approach, i.e., $g_{22}\approx b/(\ln
\xi/X)$. For instance, the data with $\xi\gtrsim 10$ are consistent
with this behavior (the fit gives $b=-0.482(4)$ with
$\chi^2/{\rm DOF}\approx 1.1$).  At $\sigma=0.0576$ the $1/(\ln \xi)^2$
extrapolation of the data satisfying 
$7\lesssim \xi\lesssim 16$ gives $g_{22}^*=-0.008(6)$
with $\chi^2/{\rm DOF}\approx 1.3$.  The data of $g_{22}$ at $\sigma\approx
0.30$ are larger, but this can be explained by crossover effects, since this
value of $\sigma$ is quite close to the critical point along the N line, where
the critical behavior may change, see Sec.~\ref{HTresnline}.

Overall the results for $g_{22}$ suggest a nonuniversal critical value.

\subsection{Critical behavior of the overlap correlations}
\label{overlapcb}

\begin{figure}[tb]
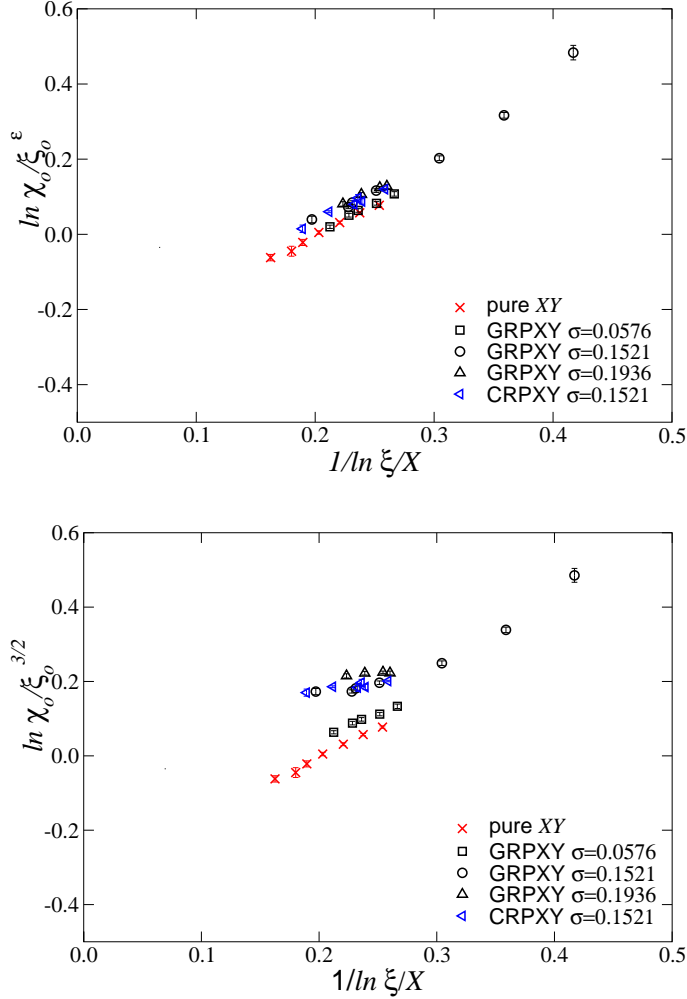

\begin{center}
\psfig{width=9truecm,angle=0,file=etao2.eps}
\hbox to \hsize {\hss}
\psfig{width=9truecm,angle=0,file=etao.eps}
\caption{ MC estimates of $\chi_o/\xi_o^{\varepsilon(\sigma)}$ (above),
where $\varepsilon(\sigma)=2-\eta_o(\sigma)$,
and $\eta_o(\sigma)$ is given by 
(\ref{etaoG}) and (\ref{etaoC}), and of 
$\chi_o/\xi_o^{2- \eta_o}$ (below), where we take 
the pure XY exponent $\eta_o=1/2$.
}
\label{etao}
\end{center}
\end{figure}

We now discuss the critical behavior of overlap correlations, 
cf. (\ref{overlap}), which are 
the appropriate quantities to understand the role of disorder.
We consider the critical behavior of the overlap susceptibility
which is expected to behave as $\chi_o\sim \xi_o^{2-\eta_o}$.  In the case of
the pure XY model we have $\eta_o=2\eta=1/2$.  In \cite{APV-09} it was
noted that the following relations
\begin{eqnarray}
&2 \eta - \eta_o  \approx 
\displaystyle{\sigma\over \pi} \quad & {\rm for} \quad {\rm GRPXY},
\label{testdiffG}\\
&2 \eta - \eta_o \approx \displaystyle{\sigma + {1\over 2} \sigma^2\over \pi} 
\quad & {\rm for} \quad {\rm CRPXY}
\label{testdiffC}
\end{eqnarray}
approximately hold in the whole QLRO phase (within the small statistical
errors), even very close to the KT transition, as long as 
$\sigma$ is not to large (in practice $\sigma$ should not be close
to $\sigma_M$, where $M$ is the Nishimori point defined in Fig.~\ref{phdia}). 
This would suggest that they
may remain valid up to the transition. Given the strong numerical evidence
that the exponent $\eta$ associated with the magnetic correlation is
$\eta=1/4$, see Sec.~\ref{ktmagncorr}, the
above relations imply that $\eta_o$ varies along
the paramagnetic-QLRO transition line approximately as
\begin{eqnarray}
\eta_o\approx {1\over 2} - {\sigma\over \pi} 
\quad & {\rm for} \quad {\rm GRPXY},\label{etaoG}\\
\eta_o\approx {1\over 2} - {\sigma+\sigma^2/2\over \pi}
\quad & {\rm for} \quad {\rm CRPXY}
\label{etaoC},
\end{eqnarray}
at least for sufficiently small values of $\sigma$.  We wish now to verify if
the high-temperature data are consistent with these predictions. In
Fig.~\ref{etao} we plot $\chi_o/\xi^{2-\eta_o}$ versus $1/\ln(\xi/X)$.  The
scaling is reasonable. We also report $\chi_o/\xi^{2-\eta_o}$, fixing 
$\eta_o$  to the pure
XY value $\eta_o=1/2$. Again the ratio is consistent with a limiting finite
value. However, if $\chi_o$ behaves as in the pure XY model, 
we would expect a $\sigma$-independent slope, see \ref{AppRGsigmazero},
which is not supported by the data.

\begin{figure}[tpb]
\begin{center}
\psfig{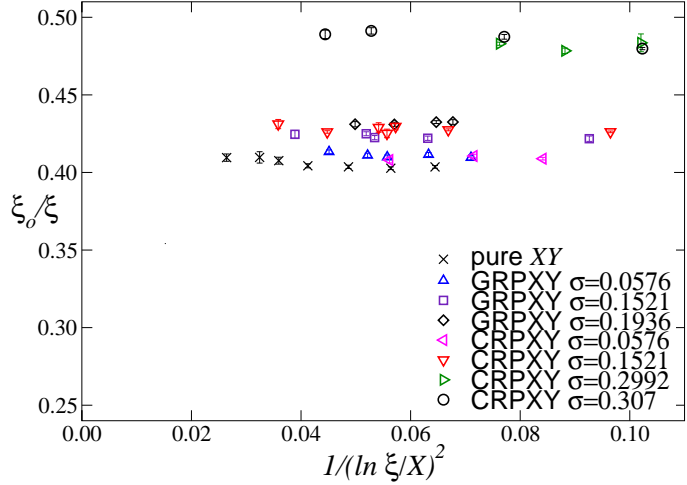}
\caption{
  The ratio $\xi_o/\xi$ versus $1/(\ln \xi/X)^2$ for the models
  considered. }
\label{raxi}
\end{center}
\end{figure}

We now consider the ratio $\xi_o/\xi$ between the second-moment correlation
lengths obtained from the overlap and spin correlation functions, cf.
~(\ref{smc}).\footnote{In a Gaussian theory without disorder, in which 
  the magnetic correlation function is given by 
  $\widetilde{G}(p)=(p^2+m^2)^{-1}$, one can easily find that
  $\xi_o/\xi=\sqrt{1/6}=0.408248...$} In order to estimate this ratio in the
case of the pure XY model, we performed MC simulations (using the
cluster algorithm) in the range $0.93\le \beta\le 1.033$ corresponding to
$12\lesssim \xi \lesssim 110$.  Taking into account the logarithmic scaling
corrections, i.e. fitting the XY-model data satisfying $\xi\gtrsim 32$ to $a
+ b/(\ln \xi/X)^2$ with $X=0.233$, we obtain the estimate
$\xi_o/\xi=0.417(4)$. In Fig.~\ref{raxi} we show the results for several
values of $\sigma$.  They are all consistent with a finite critical value,
confirming that the paramagnetic-QLRO transitions are characterized by a
single diverging length. The results can be extrapolated by assuming
$\xi_o/\xi = a+b/(\ln \xi/X)^2$ for $\xi\to\infty$. We obtain
$\xi_o/\xi=0.417(5),\,0.428(5),\,0.425(7),\,0.425(3)$ for the GRPXY model at
$\sigma=0.0576,\,0.1521,\, 0.1936$ and the CRPXY model at $\sigma=0.1521$,
respectively.  A larger result is found for the CRPXY model at $\sigma\approx
0.299$, 0.307: $\xi_o/\xi\approx 0.49$.

These results indicate that the ratio $\xi_o/\xi$ varies along the transition
line, although it changes very weakly for small values of $\sigma$. Again,
this is consistent with the observation that disorder-related quantities, like
$\eta_o$ and $g_{22}$, depend on $\sigma$.

\section{Critical behavior along the N line in the CRPXY model}
\label{HTresnline}

We now consider the critical behavior along the N line $T=\sigma$ 
in the CRPXY model, approaching the transition point from the paramagnetic
phase. We recall that along the N line the magnetic and overlap correlation
functions are equal, so that $\eta_o=\eta$ and $\xi_o=\xi$ exactly. We performed
several MC simulations along the N line, in the range $1.5\le \beta \le 2.4$,
corresponding to $2\lesssim \xi \lesssim 28$, and considered 
large lattice sizes, in order to obtain infinite-volume results.

Our MC estimates of the magnetic correlation length $\xi$ are consistent with
an exponential increase, i.e. with a behavior of the form ${\ln \xi} \sim
t^{-1/2}$ with $t=T/T_M-1$, see Fig.~\ref{xi}.  A linear fit to 
\begin{equation}
{\ln \xi} = a t^{-1/2} + b
\label{fitmnp}
\end{equation}
of the data satisfying $\xi\gtrsim 5$ gives the estimate
\begin{equation}
T_M=\sigma_M=0.307(2), 
\label{tmsm}
\end{equation}
with $\chi^2/{\rm DOF}\lesssim 1.0$. We also mention that alternative fits to
$\xi = a t^{-b}$ and to ${\ln \xi} = a t^{-1} + b$ give rise to significantly
larger $\chi^2$.

In order to estimate the exponent $\eta$, we fit $\chi$ and
$\xi$ to $\chi = c \xi^{2-\eta}$. Considering the MC results
satisfying $\xi\gtrsim \xi_{\rm min} = 5$, we find $\eta=0.246(4)$ 
with $\chi^2/{\rm DOF}\approx 1.0$. If we increase $\xi_{\rm min}$,
$\eta$ slightly decreases, 
but it is always compatible with $\eta=1/4$.  These results suggest that
$\eta=1/4$ also along the N line.  

\begin{figure}[tpb]
\begin{center}
\psfig{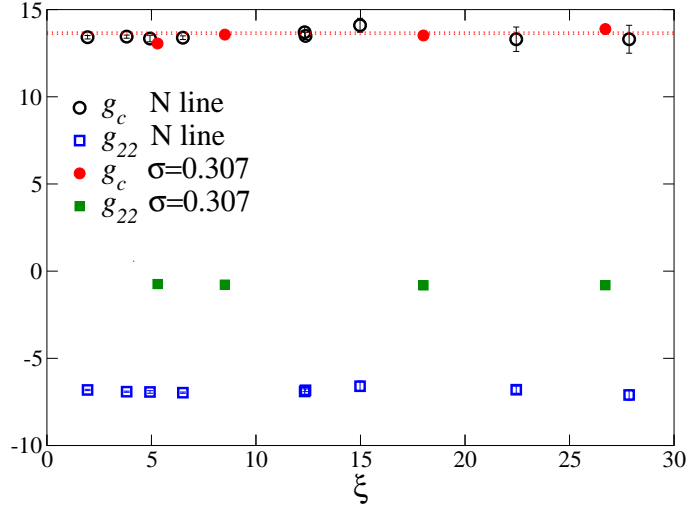}
\caption{ 
  Estimates of $g_c$ and $g_{22}$ along the N line and at $\sigma=0.307$.  The
  dotted lines indicate the estimate~\cite{BNNPSW-01} 
  $g_c^*=g_{4,XY}^*=13.65(6)$ for
  the pure XY model.  }
\label{g4g22nline}
\end{center}
\end{figure}

Fig.~\ref{g4g22nline} shows the estimates of  $g_c$.  The critical
limit of $g_c$ is consistent with the results for the pure XY
model and those obtained along the thermal paramagnetic-QLRO line at 
smaller values of $\sigma$, see Fig.~\ref{g4}.  
Indeed, a fit of all data of $g_c$ to
(\ref{g4beh}) gives $g_c^*=13.49(13)$ with $\chi^2/{\rm DOF}\approx 0.6$.
If we consider only the data satisfying $\xi\gtrsim 4$, 
we obtain $g_c^*=13.6(3)$.  

The above-reported results (KT behavior of $\xi$, $\eta=1/4$, and
$g_c^*\approx g_{4,XY}^*$) suggest that the magnetic correlations behave as in
the pure XY model. There is, however, a result which contradicts this
hypothesis. As we discussed in Sec.~\ref{XYpuro}, the rate of approach of
$\chi \xi^{-7/4}$ to its limiting value, should be universal. As can be seen
from Fig.~\ref{eta}, this is not the case: the slope of the data along the N
line is clearly different from that predicted for the pure XY model. Thus,
even though at the Nishimori point the magnetic critical behavior is the same
as that observed along the thermal paramagnetic-QLRO transition line,
corrections are different, implying the presence of a new (probably marginal)
RG operator, which only contributes to scaling corrections in magnetic
quantities.

A better evidence for the presence
of a new, disorder-related operator is obtained by considering $g_{22}$
and $\xi_o/\xi$.
In Fig.~\ref{g4g22nline} we also report estimates of $g_{22}$ along
the N line and along the line $\sigma = 0.307$. If the estimate 
(\ref{tmsm}) is correct, the two lines intersect the critical line
at the same point, the Nishimori point.
It is quite clear from the data that
the limiting value of $g_{22}$ along the two lines is quite different.
A fit of all available data on the Nishimori line to
$g_{22}^*+b/(\ln \xi/X)^2$ gives $g_{22}^*=-7.00(5)$ 
with $\chi^2/{\rm DOF}\approx
0.9$. On the other hand, a fit of the data along the line 
at fixed $\sigma=0.307$ gives $g_{22}^*\simeq -0.8$. 
The same phenomenon is observed for the ratio $\xi_o/\xi$. As can be seen
in Fig.~\ref{raxi}, for $\sigma=0.307$ this ratio is approximately equal to 
0.49, which is clearly different from the result that holds 
exactly along the Nishimori line, $\xi_o/\xi=1$.
The large differences of the values of these two RG invariant 
quantities along the two lines
provide compelling evidence that the Nishimori point is 
a multicritical point as in the 2D $\pm J$ Ising model \cite{HPPV-08}.

To understand this conclusion, let us review 
the basic results that apply to multicritical points.
The singular part of the free energy should obey a scaling law
\begin{eqnarray}
{\cal F}_{\rm sing}(u_1,u_2) = 
b^{-d} {\cal F}_{\rm sing}(b^{y_1} u_1,b^{y_2} u_2),
\end{eqnarray}
where $u_1$ and $u_2$ are two relevant scaling fields.  They can be inferred
by using the following facts: (i) the transition line at $M$ must be
parallel to the $T$ axis, since it has been proved~\cite{ON-93} that
$\sigma_M$ is an upper bound for the values of $\sigma$ where QLRO can exist;
(ii) the condition $T=\sigma$ at the N line is RG invariant.  
We therefore have
\begin{equation}
u_1 = \sigma-\sigma_M + ...
\label{u1scalfields}
\end{equation}
where the dots indicate nonlinear corrections, which are quadratic in
$\Delta\sigma\equiv \sigma-\sigma_M$ and $\Delta T\equiv T-T_M$, so that the
line $u_1=0$ runs parallel to the $T$ axis at $M$.  Moreover, we choose
\begin{equation}
u_2=T-\sigma,
\label{scalfields}
\end{equation}  
so that the N line corresponds to $u_2=0$.  

Close to the multicritical point,
any RG invariant quantity, such as $g_{22}$, 
is expected to behave as
\begin{equation}
R = f_R(u_1 u_2^{-y_1/y_2}).
\label{rgincmcp}
\end{equation}
Now, the N-line corresponds to $u_2=0$, so that a RG invariant quantity
converges to $f_R(\infty)$. On the other hand, the line $\sigma=\sigma_M$
corresponds to $u_1 = 0$, so that a RG invariant quantity converges to
$f_R(0)$ which is generically expected to be different from
$f_R(\infty)$.  Thus, if the Nishimori point is multicritical, we
expect RG invariant quantities to have a different critical value along the
two lines. This is exactly what we observe for $g_{22}$ and $\xi_o/\xi$. Thus,
in view of the numerical results we conclude that the Nishimori point is a
multicritical point. 

It is interesting to note that the multicritical behavior is not observed in 
the magnetic sector. For instance, $g_c^*$ along the paramagnetic-QLRO 
line is equal to its XY value $g_{4,XY}^*$. The same result holds along the 
Nishimori line. In terms of the scaling function $f_{g_c}$ defined in 
(\ref{rgincmcp}) these results imply
\begin{equation}
f_{g_c}(0) = f_{g_c}(\infty) = g_{4,XY}^*~.
\end{equation}
It is then natural to conjecture that $g_c^* = g_{4,XY}^*$ along any line that
intersects the Nishimori point, i.e. that $f_{g_c}(x) = g_{4,XY}^*$ for any
$x$. The absence of multicritical behavior in the magnetic sector is also
supported by the fact that $\xi$ always shows a KT behavior and that the
magnetic exponent $\eta$ at the Nishimori point is equal to the pure-XY value
1/4.

The results we have presented should apply to generic RPXY model. In all cases
we expect a multicritical point along the paramagnetic-QLRO transition line.
It follows from universality that, at the multicritical point,
the magnetic and the overlap correlation functions have the 
same critical behavior---hence, we have $\eta=\eta_o$---though 
they may not be necessarily equal
as is the case for the CRPXY model. Note that this point is not 
expected in general to coincide with that in which the
tangent to the transition line is parallel to the $T$ axis.

\section{Glassy critical behavior at $T=0$} 
\label{glassybeh}

In the limit $\sigma\rightarrow \infty$ the RPXY model corresponds to the
gauge-glass model in which the phase shifts are uniformly distributed.  This
model has been extensively studied both at zero and at finite temperature
\cite{FTY-91,ES-85,HS-90,RTYF-91,Li-92,Gingras-92,%%
  DWKHG-92,RY-93,NK-93,Nishimori-94,JKC-95,HWFGY-95,BY-96,KS-97,%%
  KCRS-97,MG-98,Granato-98,KA-99,CP-99,Kim-00,AK-02,HO-02,KY-02,%%
  Katzgraber-03,HKM-03,CTH-03,NW-04,KC-05,TT-05,UKMCL-06,YBC-06,CLL-08}.
\cite{NK-93,Nishimori-94} showed that no long-range glassy order can
exist at finite temperature.  Although this result does not exclude the
possibility of a finite-temperature transition with an exotic low-temperature
glassy
phase, for example a phase characterized by glassy QLRO,
most numerical works
\cite{FTY-91,RY-93,Granato-98,AK-02,KY-02,Katzgraber-03,NW-04,KC-05,TT-05}
support a zero-temperature glassy critical behavior.  The
overlap correlation length $\xi_o$ diverges as $T^{-\nu}$ for $T\to 0$.  We
mention the estimates \cite{KY-02} $1/\nu=0.39(3)$ and \cite{NW-04}
$1/\nu=0.36(3)$ from finite-temperature MC simulations, and~\cite{AK-02}
$1/\nu=0.36(1)$ and ~\cite{TT-05} $1/\nu\approx 0.45$ from $T=0$ numerical
calculations.  Moreover, if one assumes that the ground state is nondegenerate
in the overlap variables, one obtains that at $T=0$ the finite-size overlap
susceptibility satisfies the relation $\chi_o = L^2$, so that $\eta_o = 0$.
We mention that this scenario was questioned in
\cite{Kim-00,CP-99,HO-02,HKM-03,CTH-03,UKMCL-06,YBC-06,CLL-08}, which
claimed the existence of a finite-temperature transition at $T\approx 0.2$.

A natural scenario for the phase digram of the GRPXY and CRPXY models is that
the glassy transition, which occurs for $\sigma=\infty$, is not isolated but
that it is the endpoint of a phase transition line that starts at the
paramagnetic-QLRO transition line.  In particular, if the zero-temperature
glassy transition scenario applies to the gauge-glass model, we expect a line
of $T=0$ glassy transitions for any $\sigma>\sigma_D$, see Fig.~\ref{phdia}. A
natural conjecture would be that all these transitions belong to the same
universality class.

To check this scenario we performed MC simulations of the CRPXY model at
$\sigma=2/3,\,5/9,\,1/2$, $\infty$, which are larger than $\sigma_D\le
\sigma_M\approx 0.31$.  As we shall see, the results clearly support a glassy
$T=0$ transition in the same universality glass as that of the gauge-glass
model.

\subsection{MC simulations}
\label{MCglassy}

We performed MC simulations of the CRPXY model on square $L\times L$ lattices
with periodic boundary conditions.  Most of the results we shall present refer 
to runs with $\sigma=2/3$. In this case we considered $L=20$, 30,
40, 60, 80 and temperatures between $T=2/3$ (at the Nishimori line) and
$T=0.1$ (for $L=80$ we considered $0.22 \le T \le 2/3$). We averaged over a
relatively large number $N_s$ of samples: $N_s=6000$, 9000, 7000, 3000, and
2000 samples for $L=20$, 30, 40, 60 and 80, respectively. We used the MC
algorithm discussed in \ref{AppMC} combined with the parallel-tempering
method \cite{raex,par-temp}. 
Moreover, to check the universality of the transitions, we
also performed parallel-tempering MC simulations 
for $\sigma=5/9$ and lattice sizes $L=60,\,70$ (5000 and 1000
disorder samples, respectively), $\sigma=1/2$ and $L=70$ (1000 samples),
and $\sigma=\infty$ and $L=20$, 30, 40, 60 (5000, 5000, 2000, 2000 samples, 
respectively).
The points in the $T$-$\sigma$ plane where we collected MC data are shown in
Fig.~\ref{tc}.

\begin{figure}[tpb]
\begin{center}
\psfig{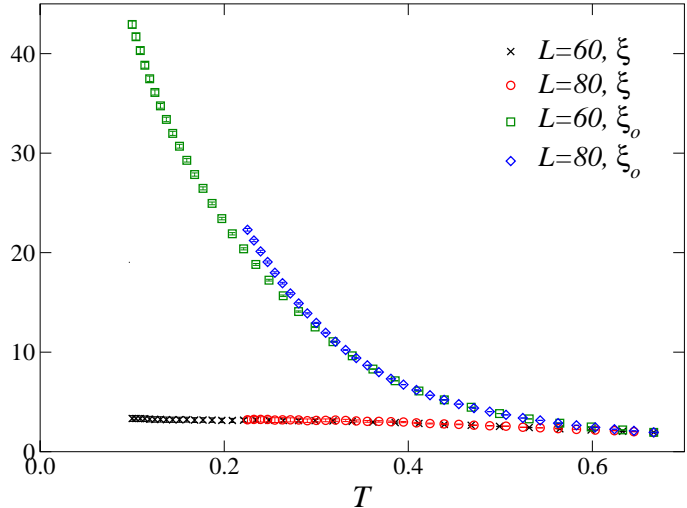}
\caption{ 
  MC estimates of the correlation lengths $\xi$ and $\xi_o$ for the 
  CRPXY model at $\sigma=2/3$.  }
\label{xislt}
\end{center}
\end{figure}

At the glassy transition the critical modes are those related to the overlap
variables, while the magnetic ones are noncritical.  This is clearly shown in
Fig.~\ref{xislt}, which shows $\xi$ and $\xi_o$ for $\sigma=2/3$.  
The overlap correlation
length $\xi_o$ increases steadily with decreasing the temperature, while the
magnetic correlation length $\xi$ freezes at sufficiently low temperatures at a
value $\xi\approx 3.3$. Therefore, the 
critical temperature and exponents must be determined from
quantities related to the overlap correlation functions.

\subsection{Evidence for a $T=0$ glassy transition at $\sigma=2/3$}
\label{tc2o3}

\begin{figure}[tpb]
\begin{center}
\psfig{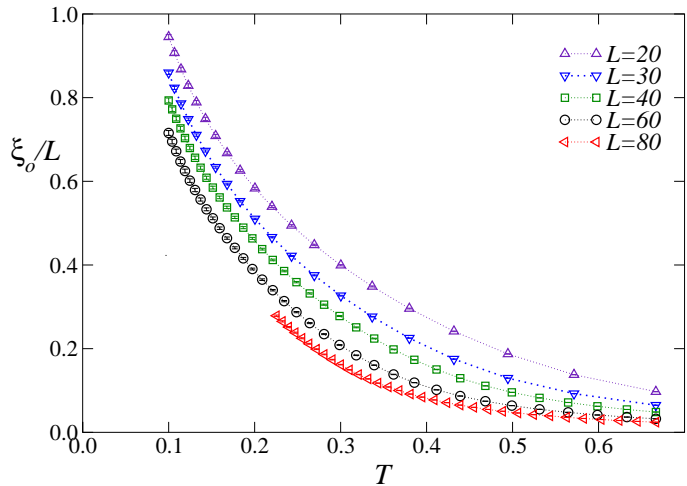}
\caption{ 
MC estimates of the ratio $R_{\xi_o}\equiv \xi_o/L$
for the CRPXY model at  $\sigma=2/3$.  
}
\label{xioovl}
\end{center}
\end{figure}

\begin{table}
\footnotesize
\begin{center}
\begin{tabular}{clcl}
\hline\hline
$L_{\rm min}$ & \multicolumn{1}{c}{$T_{\rm max}$} & $\chi^2$/DOF & 
\multicolumn{1}{c}{$T_c$} \\
\hline
  20    &   0.6  &  169/157  &  0.018(1)   \\
  20    &   0.5  &   99/141  &  0.009(1)   \\
  20    &   0.4  &   67/119  &  0.010(2)   \\
  20    &   0.3  &   30/92   &  0.010(3)   \\
\hline
  30    &   0.6  &  137/138  &  0.017(1)   \\
  30    &   0.5  &   66/123  &  0.008(2)   \\
  30    &   0.4  &   49/103  &  0.007(3)   \\
  30    &   0.3  &   21/79   &  0.005(4)   \\
\hline
  40    &   0.6  &  106/119  &  0.017(2)   \\
  40    &   0.5  &   42/105  &  0.007(2)   \\
  40    &   0.4  &   31/87   &  0.007(3)   \\
  40    &   0.3  &   17/66   &  0.007(5)   \\
\hline\hline
\end{tabular}
\end{center}
\caption{Estimates of $T_c$ obtained by fitting $R_\xi$ to 
(\protect\ref{fitRxi}) with $n=6$.
DOF is the number of degrees of freedom of the fit. }
\label{Tc-glass}
\end{table}

In order to determine the critical temperature, we analyze 
$R_{\xi_o}\equiv \xi_o/L$. The results, shown in Fig.~\ref{xioovl}, 
show no evidence of a crossing point in the range of values of $T$ of the data,
$T\ge 0.1$, and thus provide the bound $T_c<0.1$ for the critical
temperature $T_c$. A more precise determination of $T_c$ can be
obtained by a finite-size scaling (FSS) analysis.  We fit the data to
\begin{equation}
R_{\xi_o} = P_n[(T-T_c) L^{1/\nu}],
 \label{fitRxi}
\end{equation}
keeping $T_c$ and $\nu$ as free parameters. Here $P_n(x)$ is a polynomial in
$x$ of order $n$. The order $n$ is fixed by looking at the $\chi^2$ of the fit.
For each $n$ we determine the goodness $\chi^2(n)$ of the fit. Then, we fix
$n$ such that $\chi^2(n)$ is not significantly different from $\chi^2(n+1)$.
The results we report correspond to $n=6$. To identify the role of the
corrections to scaling we repeat the fit several times.  Each time we fix two
parameters $T_{\rm max}$ and $L_{\rm min}$ and we only include the data which
correspond to lattices satisfying the conditions $T \le T_{\rm max}$ and $L\ge
L_{\rm min}$.

In Table \ref{Tc-glass} we report the estimates of $T_c$ for several values of
$T_{\rm max}$ and $L_{\rm min}$. We obtain estimates of $T_c$ which are quite
small and satisfy the upper bound
\begin{equation}
   T_c \lesssim 0.01~.
\end{equation}
Since our data satisfy $T\ge 0.1$, this estimate
allows us to conclude that our results are fully consistent with 
a zero-temperature transition. From now on, we always assume $T_c = 0$.

\subsection{The critical exponent $\nu$}
\label{critexpnu}

\begin{table}
\footnotesize
\begin{center}
\begin{tabular}{clcl}
\hline\hline
$L_{\rm min}$ & \multicolumn{1}{c}{$T_{\rm max}$} & $\chi^2$/DOF & 
\multicolumn{1}{c}{$\nu$} \\
\hline
  20   &  0.4    &  100/120   &        2.465(6)  \\
  20   &  0.3    &   44/93    &        2.496(10) \\
  20   &  0.25   &   24/77    &        2.528(14) \\
  20   &  0.2    &   17/55    &        2.547(22) \\
  20   &  0.16   &   14/39    &        2.548(31) \\
\hline
  30   &  0.4    &   55/104    &       2.446(6)  \\
  30   &  0.3    &   23/80     &       2.464(13) \\
  30   &  0.25   &   13/65     &       2.489(20) \\
  30   &  0.2    &   10/46     &       2.492(30) \\
  30   &  0.16   &   9/32      &       2.488(42) \\
\hline
  40   &  0.4     &  36/88      &      2.432(7)  \\
  40   &  0.3     &  19/67      &      2.451(15) \\
  40   &  0.25    &  12/53      &      2.480(26) \\
  40   &  0.2     &   8/37      &      2.490(38) \\
  40   &  0.16    &   9/25      &      2.482(53) \\
\hline\hline
\end{tabular}
\end{center}
\caption{Estimates of $\nu$ obtained by fitting $R_{\xi_o}$ to 
(\ref{fitRxi}) with $T_c = 0$ and $n=6$. 
DOF is the number of degrees of freedom of the fit. }
\label{nu-glass}
\end{table}

In order to determine the critical exponent $\nu$ related to the divergence of
the correlation length $\xi_o$, we repeat the fit (\ref{fitRxi}) 
at $\sigma=2/3$ setting $T_c = 0$. The results are reported in Table
~\ref{nu-glass}.  They slightly increase as $T_{\rm max}$ or $L_{\rm min}$ is
lowered, but these changes are small compared to the statistical errors. 

In fit (\ref{fitRxi}) we made two approximations. First, we neglected the
nonanalytic scaling corrections, which decrease as $L^{-\omega}$.  The results
indicate that these corrections are small: at fixed $T_{\rm max}<0.25$ the
estimates of $\nu$ obtained setting $L_{\min} = 30$ and $L_{\min} = 40$ differ
by much less than the statistical errors. Second, we approximated the thermal
nonlinear scaling field $u_T$ by $u_T\approx T$, neglecting the {\em analytic}
corrections (see \cite{HPV-08} for an extensive discussion of this type
of corrections).  To understand their quantitative role, we performed fits to
\begin{equation}
R_\xi = P_n(u_T L^{1/\nu}), \qquad u_T \equiv T + p T^2,
\label{fitRxi-analytic}
\end{equation}
where $p$ is a new free parameter. The results are reported in
Table~\ref{nu-glass-analytic}. Corrections are tiny and we estimate
$|p|\lesssim 0.2$, so that $|u_T-T|/T$ is at most 0.10, 0.02 for $T=0.5$, 0.1,
respectively.  The estimates of $\nu$ do not vary significantly and, for $L\ge
30$ and $T_{\rm max}\le 0.2$, are fully consistent with those obtained before.
We quote
\begin{equation}
\nu = 2.5(1)~, \qquad 1/\nu = 0.40(2)
\label{nufinale}
\end{equation}
as our final estimate.

\begin{figure}[tpb]
\begin{center}
\psfig{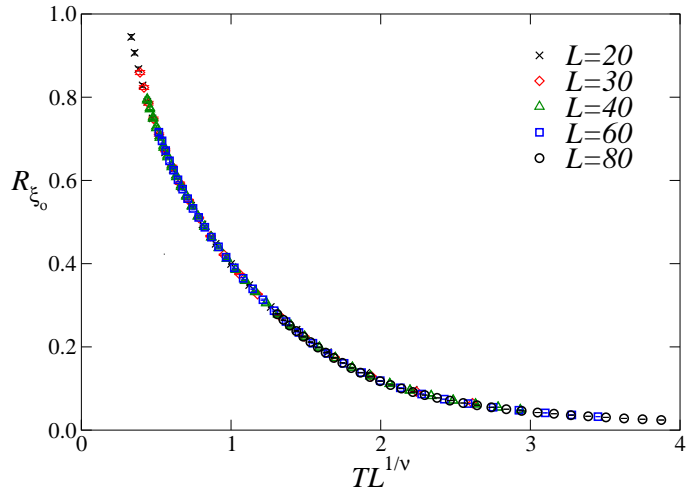}
\caption{ 
  $R_{\xi_o}\equiv \xi_o/L$ versus $TL^{1/\nu}$ for $\nu = 2.5$. 
  Data corresponding to $\sigma = 2/3$.
}
\label{FSSxi}
\end{center}
\end{figure}

To show the quality of our FSS results in Fig.~\ref{FSSxi} we plot $R_{\xi}$
versus $TL^{1/\nu}$, using the estimate (\ref{nufinale}). All data fall on top
of each other with remarkable precision.

\begin{table}
\footnotesize
\begin{center}
\begin{tabular}{clcll}
\hline\hline
$L_{\rm min}$ & \multicolumn{1}{c}{$T_{\rm max}$} & $\chi^2$/DOF & 
\multicolumn{1}{c}{$\nu$} & 
\multicolumn{1}{c}{$p$}   \\
\hline
   20  &    0.5  &     98/141    &    2.54(1)  &     $-$0.11(1) \\
   20  &    0.4  &     63/119    &    2.62(2)  &     $-$0.20(2) \\
   20  &    0.3  &     28/92     &    2.71(4)  &     $-$0.34(5) \\
   20  &    0.25 &     20/76     &    2.67(6)  &     $-$0.26(11) \\
   20  &    0.2  &     16/54     &    2.67(10) &     $-$0.29(21) \\
\hline
   30  &    0.5  &     89/123    &    2.42(2)  &      $-$0.00(2) \\
   30  &    0.4  &     47/103    &    2.54(2)  &      $-$0.12(3) \\
   30  &    0.3  &     20/79     &    2.58(5)  &      $-$0.18(7) \\
   30  &    0.25 &     13/64     &    2.54(7)  &      $-$0.11(14) \\
   30  &    0.2  &     10/45     &    2.50(12) &      $-$0.01(31) \\
\hline
   40  &    0.5  &     50/105    &    2.42(2)  &      $-$0.00(2) \\
   40  &    0.4  &     31/87     &    2.50(3)  &      $-$0.09(3) \\
   40  &    0.3  &     17/66     &    2.58(6)  &      $-$0.20(7) \\
   40  &    0.25 &     12/52     &    2.54(16) &      $-$0.12(16) \\
   40  &    0.2  &     10/36     &    2.50(13) &      $-$0.01(33) \\
\hline\hline
\end{tabular}
\end{center}
\caption{Estimates of $\nu$ and $p$ obtained by fitting $R_{\xi_o}$ to 
(\ref{fitRxi-analytic}) with $n = 6$. 
DOF is the number of degrees of freedom of the fit.}
\label{nu-glass-analytic}
\end{table}

\subsection{The critical exponent $\eta_o$}
\label{critexpeta}

\begin{figure}[tpb]
\begin{center}
\psfig{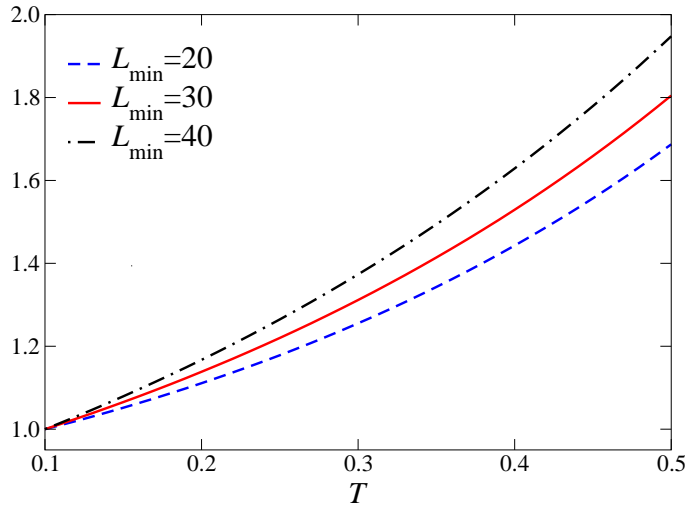}
\caption{ 
Plot of the ratio $\overline{u}_h(T)/\overline{u}_h(T=0.1)$ 
  from fits with $T_{\rm max} = 0.5$ and $L_{\rm min} =20$, 30, 40.  
}
\label{uh}
\end{center}
\end{figure}

As discussed at length in \cite{HPV-08}, the overlap susceptibility behaves
in the critical limit as
\begin{equation}
\chi_o = \overline{u}_h^2 L^{2-\eta_o} f(u_T L^{1/\nu})~.
\end{equation}
Here $u_T$ is the temperature nonlinear scaling field, while $\overline{u}_h$
is related to the external {\em overlap-magnetic} scaling field $u_h$
associated with the overlap variables by $u_h=h\overline{u}_h(T)+ O(h^2)$.  We
have already checked that the thermal scaling field $u_T$ can be effectively
approximated by $u_T = T$.  Thus, neglecting nonanalytic scaling corrections,
the data should behave as
\begin{equation}
\ln \chi_o = (2-\eta_o) \ln L + \ln \overline{u}_h(T)^2 + 
   \ln f(T L^{1/\nu})~.
\end{equation}
We now estimate $\eta_o$ from the analysis of the data at $\sigma=2/3$.  In a
first set of fits we set $\overline{u}_h = 1$ and approximate $\ln f(x)$ with
a polynomial in $x$ of order $n$, i.e., we perform fits to
\begin{equation}
\ln \chi_o = (2-\eta) \ln L + P_n(T L^{1/\nu}).
\label{fit1-eta-glass}
\end{equation}
The analysis of the $\chi^2$ of the fits indicate that $n=6$ allows us to
describe accurately the data.
\begin{table}
\footnotesize
\begin{center}
\begin{tabular}{clclcl}
\hline\hline
 &&
\multicolumn{2}{c}{Fit (\ref{fit1-eta-glass})} &
\multicolumn{2}{c}{Fit (\ref{fit2-eta-glass})}  \\
$L_{\rm min}$ & \multicolumn{1}{c}{$T_{\rm max}$} & $\chi^2$/DOF & 
\multicolumn{1}{c}{$\eta$} &   $\chi^2$/DOF &
\multicolumn{1}{c}{$\eta$}  \\
\hline
   20   &    0.5  &   11570/142  &    0.13(2)  &   338/140  &  $-$0.01(1) \\
   20   &    0.4  &   1439/120   &    0.10(2)  &    99/118  & \phm0.02(1) \\
   20   &    0.3  &    498/93    &    0.06(1)  &    39/91   & \phm0.04(3) \\
   20   &    0.25 &    182/77    &    0.06(1)  &    20/75   & \phm0.01(3) \\
   20   &    0.2  &     43/55    &    0.05(1)  &    12/53   &  $-$0.04(8) \\
\hline
   30   &    0.5  &   6592/124   &    0.17(2)  &   263/122  &  $-$0.03(2) \\
   30   &    0.4  &   1096/104   &    0.11(2)  &    78/102  & \phm0.01(2) \\
   30   &    0.3  &    330/80    &    0.07(1)  &    35/78   & \phm0.04(3) \\
   30   &    0.25 &     89/65    &    0.05(1)  &    18/63   & \phm0.00(4) \\
   30   &    0.2  &     28/46    &    0.05(2)  &    11/44   &  $-$0.06(10) \\
\hline
   40   &    0.5  &   4237/106   &    0.18(2)  &   177/104  &  $-$0.05(2) \\
   40   &    0.4  &   1096/88    &    0.11(2)  &    40/86   &  $-$0.03(2) \\
   40   &    0.3  &    294/67    &    0.07(1)  &    22/65   & \phm0.02(4) \\
   40   &    0.25 &    63/53     &    0.05(1)  &     9/51   &  $-$0.02(6) \\
   40   &    0.2  &    17/37     &    0.04(1)  &     2/35   &  $-$0.11(12) \\
\hline\hline
\end{tabular}
\end{center}
\caption{Estimates of $\eta$. 
  On the left we report the results of the fits to (\ref{fit1-eta-glass})
  with $n=6$, on the right those to (\ref{fit2-eta-glass})
  with $n=6$ and $m = 2$. In both cases 
  we fix $\nu = 2.5(1)$. The reported errors are the sum of the statistical
  error and of the variation of the estimate of $\eta$ 
  as $\nu$ changes by one error bar.
  DOF is the number of degrees of freedom of the fit.}
\label{eta-glass}
\end{table}
We fix $\nu$ to the estimate (\ref{nufinale}) to avoid an additional nonlinear
parameter in the fit.  The results are reported in Table~\ref{eta-glass}. We
observe a significant change of the estimates as $T_{\rm max}$ decreases;
moreover, the quality of the fit is quite poor.  This can be explained by the
presence of sizeable analytic corrections, which means that 
$\overline{u}_h$ is poorly approximated by a 
$\overline{u}_h=1$ in our range of temperatures. The same phenomenon
occurs in the
three-dimensional Ising spin glass \cite{HPV-08}, where the analytic
corrections cannot be neglected in the analysis of the overlap
susceptibility. We thus perform a second set of fits in which we take into
account the magnetic nonlinear scaling field. If we approximate $\ln
\overline{u}_h^2$ with a polynomial of order $m$, we end up with the fitting
form
\begin{equation}
\ln \chi_o = (2-\eta) \ln L + P_n(T L^{1/\nu}) + Q_m(T)~,
\label{fit2-eta-glass}
\end{equation}
where we assume $Q_m(0) = 0$. In the following we take $m=2$ and again fix
$\nu$ to the estimate (\ref{nufinale}). The results are reported in
Table~\ref{eta-glass}. The quality of the fit is now significantly better,
indicating that the analytic corrections are important. The
scaling function $\overline{u}_h$ is reported in Fig.~\ref{uh} and indeed it
varies significantly in the range of values of $T$ we are considering. The
estimates of $\eta_o$ do not show any systematic variation with $T_{\rm max}$
and are always consistent, within errors, with $\eta_o = 0$.
Quantitatively, our data allow us to set the upper bound
\begin{eqnarray}
|\eta_o| \le 0.05.
\end{eqnarray}

\subsection{Results for the gauge-glass model} \label{gaugeglass}

In order to check universality we also performed runs at $\sigma=\infty$,
although in this case we considered smaller lattices and the errors are 
significantly larger (partly because of the smaller number of samples,
partly because of larger sample-to-sample fluctuations).
The data were analyzed as we 
did in the $\sigma=2/3$ case. First, we determined the critical temperature
$T_c$. A fit of $\xi_{o}/L$ to (\ref{fitRxi}) gives rather small 
estimates of $T_c$. For $L_{\rm min} = 20$ we obtain 
$T_c = 0.030(2)$ [0.020(3)] for $T_{\rm max} = 0.4$ (resp. 0.3). Thus,
we can conclude that $T_c\lesssim 0.02$, which is clearly
consistent with $T_c = 0$, given that our data belong
to the range $T\ge 0.1$. The claim that $T_c\approx 0.2$ is not consistent
with our MC data. 

Then, we determined $\nu$ by assuming $T_c = 0$. The results of the 
fits to (\ref{fitRxi}) show a significant dependence on $T_{\rm max}$.
For $L_{\rm min} = 20$, $\nu$ varies between 2.50(1) and 2.80(4) as 
$T_{\max}$ varies between 0.4 and 0.16. If analytic scaling corrections
are included, i.e. we fit the data to (\ref{fitRxi-analytic}), we observe
a significantly smaller dependence on $T_{\max}$, but, on the 
other hand, a rather large dependence on $L_{\rm min}$, with 
rapidly increasing error bars as $L_{\rm min}$ increases. This is 
probably due to the fact that we have a somewhat large statistical error
on the results with the largest value of $L$, $L=60$. The estimates 
of $\nu$ vary between 2.8 and 3.7 if we take $L_{\rm min} = 20$, 30 and 
$0.2\le T_{\rm max} \le 0.5$ and thus give the final result
$\nu = 3.3(5)$. This result is somewhat larger than the estimate 
(\ref{nufinale}), but certainly not inconsistent. 
It supports --- very weakly, though---universality. A better
check is presented below.

\subsection{The quartic coupling $g_{o}$ and universality}
\label{g4odata}

\begin{figure}[tpb]
\begin{center}
\psfig{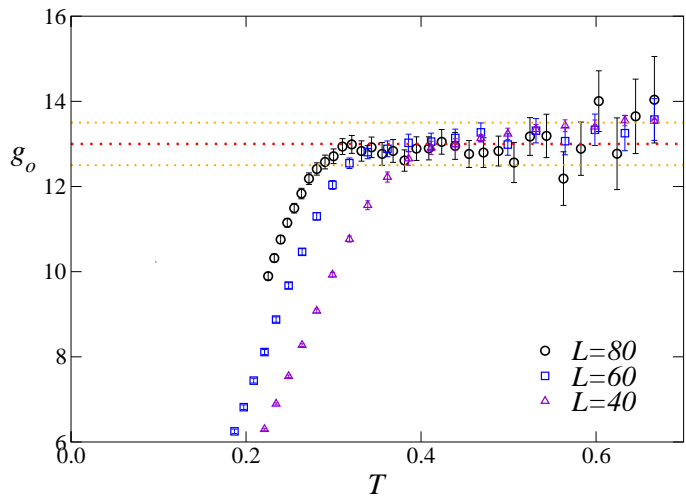}
\caption{ 
  MC estimates of $g_{o}$ vs $T$ at $\sigma=2/3$ for $L=40,\,60,\,80$.  
  The dotted lines correspond to the infinite-volume critical ($T=0$) 
  estimate $g_{o}^*=13.0(5)$.  }
\label{g4o}
\end{center}
\end{figure}

We computed the overlap quartic coupling $g_{o}$ defined in
(\ref{g4odef}).  MC results at $\sigma=2/3$ are shown in Fig.~\ref{g4o}.
The infinite-volume limit, within our statistical accuracy, is apparently
reached when $L/\xi_o \gtrsim 7$, corresponding to $T\gtrsim 0.3$ for our
largest lattices $L=60,\,80$.  The infinite-volume results are quite stable
with respect to $T$, so that we can reliably estimate the critical ($T=0$)
value $g_{o}^*$.  We obtain
\begin{equation} 
g_{o}^*=13.0(5).
\end{equation}

\begin{figure}[tpb]
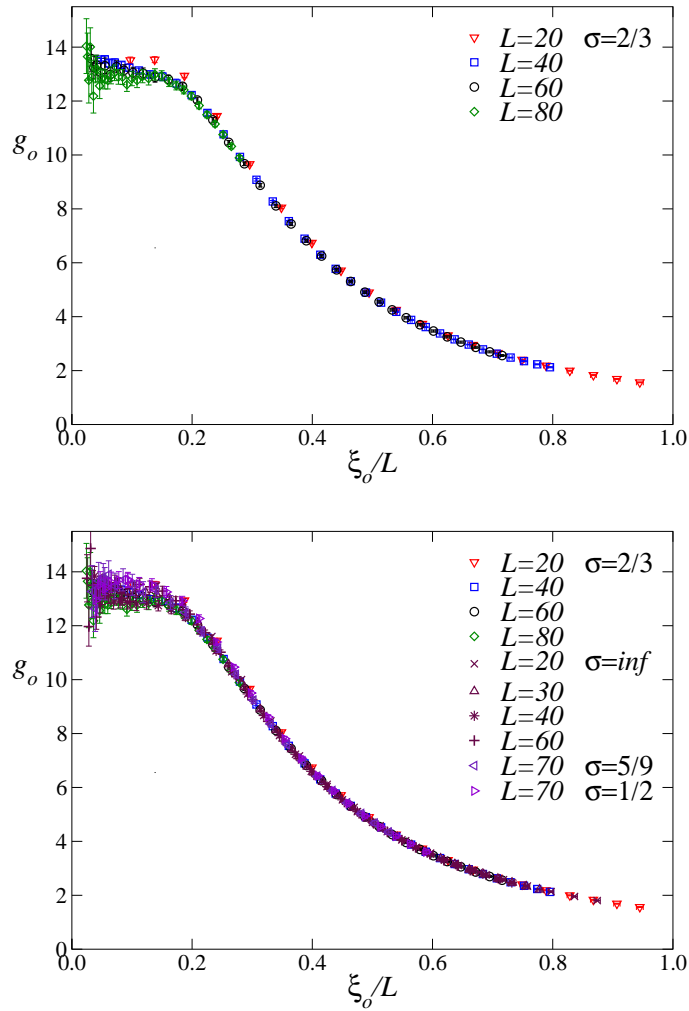

\begin{center}
\psfig{width=9truecm,angle=0,file=g4orxio.eps}
\hbox to \hsize {\hss}
\psfig{width=9truecm,angle=0,file=g4orxio2.eps}
\caption{ 
  $g_{o}$ vs $R_{\xi_o}\equiv \xi_o/L$: data at $\sigma=2/3$ 
  for various lattice sizes $L=20,40,60,80$ (above), and 
  including (below) also data for other values of $\sigma$:
  $\sigma=5/9,\,1/2,\,\infty$.}
\label{g4ovsrxi}
\end{center}
\end{figure}

According to standard RG arguments, $g_{o}$ has a universal FSS limit as a
function of $R_{\xi_o}\equiv \xi_o/L$, that is
\begin{equation}
  g_o(T,L) = f(R_{\xi_o}),  
\end{equation}
where the function $f(x)$ is universal and satisfies $f(0) = g^*_o$.  This
scaling behavior is nicely supported by the data at $\sigma=2/3$ for various
lattice sizes, see Fig.~\ref{g4ovsrxi}.  Universality can be checked by also
considering the results for $\sigma=5/9$, $\sigma=1/2$ and $\sigma=\infty$.
Clearly, all points fall on top of each other.  Note that here there are no
free parameters to fiddle with and thus this comparison provides strong
support to the hypothesis that all these models belong to the same
universality class.  Given the very good evidence we have that the model with
$\sigma=2/3$ undergoes a $T=0$ glassy transitions, this result further
confirms (and provides stronger evidence than that given in the previous
paragraph) that the gauge-glass model does not have a finite-temperature
exotic glassy transition.

\subsection{Behavior of the magnetic correlation functions}
\label{g4g22magn}

Let us now consider the magnetic quantities. The magnetic correlation length
$\xi$ is zero in the gauge-glass model, see \ref{App-magn}, and increases
as one approaches the QLRO region.  In particular, at $T=0.159$, which is
below the critical temperature $T_M\approx 0.31$ along the Nishimori line, we
obtain $\xi=3.3(1),\,6.7(4),\,9.8(3)$ at $\sigma=2/3,\,5/9,\,1/2$,
respectively.  They are roughly consistent with a behavior like $\ln \xi\sim
(\sigma-\sigma_c)^{-\kappa}$ assuming $\sigma_c\approx \sigma_M
\approx 0.30$, i.e., with a KT-like behavior along the transition
line that connects the Nishimori critical point $M$, see Fig.~\ref{phdia}, and
the $T=0$ transition point at $\sigma=\sigma_D$, which is expected to run
almost parallel to the $T$ axis.  Note, however, that while our data suggest a
power-law divergence of $\ln\xi$ (therefore, $\xi$ has an exponential
divergence), they are not sufficiently precise to allow us to estimate the
power $\kappa$. The KT value $\kappa=1/2$ is consistent with the data, but
$\kappa=1$ would be equally reasonable.

It is also interesting to discuss the behavior of the quartic couplings $g_c$,
$g_4$, and $g_{22}$ defined from the magnetic correlation functions in
(\ref{g4def})-(\ref{gcdef}). In \ref{App-magn}, assuming
universality, we predict that, in the critical limit, $g_{4}$ and
$g_{22}$ should diverge as $\xi_o^2$, while $g_c \xi_o^{-2}$ should go to
zero.

\begin{figure}[tpb]
\begin{center}
\psfig{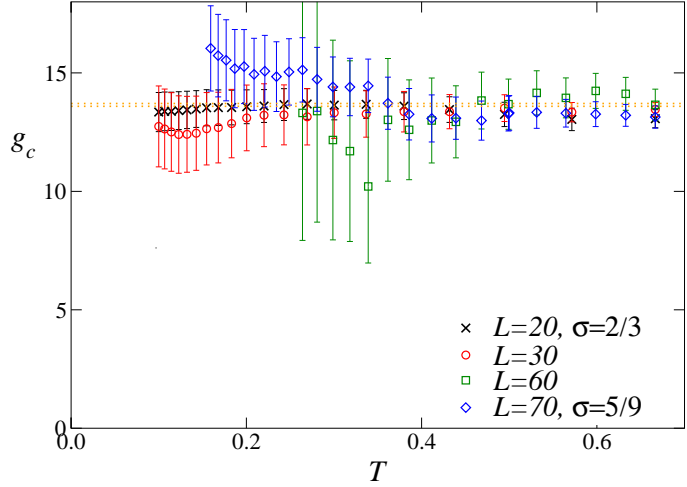}
\caption{ 
  MC estimates of the quartic couplings $g_c$. 
  The dotted line
  corresponds to the XY value $g_c=g_{4}=13.65(6)$. }
\label{g4lt}
\end{center}
\end{figure}

Numerical estimates of $g_c$ are shown in Fig.~\ref{g4lt}.  The results are
clearly consistent with a finite $T=0$ limit. Note that the estimates obtained
for $\sigma=2/3$, 5/9, and 1/2 are close to the XY value
$g_{4,XY}^*=13.65(6)$; actually, they are consistent within errors, even at
small $T$, below $T_M\approx 0.31$.  These results are suggestive of a KT
behavior of the magnetic correlation functions also along the disorder
paramagnetic-QLRO transition line from $M$ to $D$, see Fig.~\ref{phdia}.
Indeed, for $\sigma = 2/3$, 5/9, 1/2 we have $\xi \approx 3$, 7, 10, so that
along these lines one should be able to observe the critical behavior that
arises when one approaches the paramagnetic-QLRO transition line at a point
with $T<T_M$.  In other words, these results imply that the critical limit of
$g_c(\sigma,T)$ at fixed $T<T_M$ along the paramagnetic-QLRO transition line
is consistent with the KT value. This fact provides some evidence that also
along the disorder-driven transition line magnetic correlation functions
behave as in the pure XY model.  Of course, as $\sigma$ increases (thus, the
magnetic correlation length $\xi$ decreases), $g_c$ changes significantly and,
for $\sigma=\infty$, $g_c$ is infinite for any $T$ and $L$.

\begin{figure}[tpb]
\begin{center}
\psfig{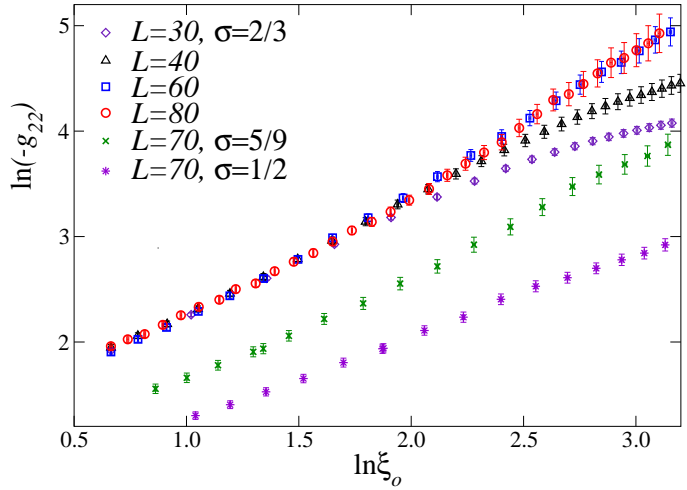}
\caption{ 
 Plot of $\ln(-g_{22})$ vs $\ln \xi_o$ at $\sigma=2/3,\,5/9,\,1/2$.
}
\label{g22vsxio}
\end{center}
\end{figure}

The couplings $g_{22}$ and $g_4$ are instead expected to diverge as $\xi_o^2$.
In Fig.~\ref{g22vsxio} we report $g_{22}$ for the different models. The data
are clearly diverging as $\xi\to\infty$, but the asymptotic behavior
$g_{22}\sim \xi_o^2$ is not clearly observed, likely because the values of
$\xi_o$ are not sufficiently large. Indeed, we only observe that 
$g_{22}$ behaves as $\xi_o^\kappa$ with $\kappa$ rapidly increasing 
with $\xi_o$. More precisely, if we only include data satisfying 
$\xi_o\lesssim 10$ we obtain $\kappa\approx 1$. If instead we fit the data 
with $10\lesssim \xi_o\lesssim 20$ (we have infinite-volume data only up
to $\xi_o\approx 20$) we obtain $\kappa\approx 1.5$.

\section{Conclusions} \label{conclusions}

We have studied the magnetic and glassy transitions of the square-lattice XY
model in the presence of random phase shifts and, in particular, the GRPXY and
CRPXY model defined by the distributions (\ref{GRPXYd}) and
(\ref{CRPXYd}). The latter is very useful because it allows some exact
calculations along the Nishimori line $T =\sigma$~\cite{ON-93,Nishimori-02},
where, in particular, the magnetic and overlap two-point functions are equal.
We present MC for the GRPXY and CRPXY models for several values of
the temperature and of the parameter $\sigma$ controlling the disorder, 
approaching the magnetic and glassy transition lines
from the paramagnetic phase.  We
substantially confirm the phase diagram shown in Fig.~\ref{phdia}.

Our main results are the following.

\begin{itemize}
\item[(i)] We have carefully investigated the critical behavior along the
  transition line separating the paramagnetic and QLRO phases, from the pure
  XY point $P$ to the multicritical point, which, in the CRPXY model, 
  lies on the N line and is such that the transition line runs parallel to the
  $T$ axis.  The magnetic observables show a $\sigma$-independent KT 
  behavior: the magnetic correlation length behaves as $\ln
  \xi \sim u_t^{-1/2}$, where $u_t$ is the thermal scaling field, $u_t\sim
  T-T_c(\sigma)$, and the magnetic susceptibility as $\chi\sim \xi^{7/4}$
  (corresponding to $\eta=1/4$).  Moreover, the quartic coupling $g_c$
  defined in (\ref{gcdef}) appears to be universal.
  We obtain  $g^*_c \approx 13.6$,
  which is nicely consistent with the corresponding value
  $g^*_{4,XY}=13.65(6)$ of the pure XY model~\cite{BNNPSW-01,PV-r}. We have also
  verified the universality of the leading logarithmic correction to the
  critical behavior of $\chi$. On the other hand, the critical behavior of
  disorder-related quantities, such as those related to the overlap
  correlation function, depends on $\sigma$.
\item[(ii)] In the CRPXY model, the Nishimori point $M$, see Fig.~\ref{phdia},
  is a multicritical point which divides the paramagnetic-QLRO line into two
  parts: a thermally-driven transition line (from $P$ to $M$) and a
  disorder-driven transition line (from $M$ to $D$). This result should be
  general: a multicritical point should also exist in generic RPXY models,
  although in this case it is not expected to coincide with that 
  where the transition line runs parallel to the $T$ axis.
  Such a multicritical point is characterized by the fact that, at 
  criticality, magnetic and overlap functions have the same critical
  behavior, that is $\eta=\eta_o$: in the CRPXY model the 
  two correlation functions are exactly equal (more generally,
  they are equal on the whole N line), but we do not expect this 
  property to be generic.
  It is interesting to observe that the multicritical behavior is only
  observed in the disorder-related quantities. Magnetic observables behave, as
  far as the leading behavior is concerned, as in the pure XY model: the
  correlation length shows a KT behavior, $\eta = 1/4$, and $g^*_c =
  g^*_{4,XY}$ in the whole neighborhood of the multicritical point. However,
  corrections are different from those appearing in the pure XY model,
  providing additional evidence for the presence of an additional (probably
  marginal) RG operator, which is responsible for the multicritical behavior.
\item[(iii)] Little is known about the behavior along the transition line from
  the multicritical point to $D$. 
   However, the fact that purely magnetic observables behave as in
  the pure XY model both along the thermally-driven transition line and at the
  multicritical point make us conjecture that the magnetic behavior is also
  unchanged. We have presented some very weak evidence in 
  Sec.~\ref{g4g22magn}.
\item[(iv)] We have investigated the critical behavior for large values of
  $\sigma$.  We find no evidence of a finite-temperature transition for all
  values of $\sigma$ we have investigated: the system is paramagnetic up to
  $T=0$, where a glassy transition occurs.  Morever, in all cases we verify
  universality. We can thus conjecture that the critical behavior along the
  whole line that starts in $D$, see Fig.~\ref{phdia}, is universal: for any
  $\sigma > \sigma_D$, one has the same critical behavior characterized by
  the exponents:
\begin{equation}
\nu = 2.5(1), \quad 1/\nu = 0.40(2),
\qquad |\eta_o| \le 0.05. 
\end{equation}
Our estimate of $\nu$ is consistent with earlier estimates
obtained by MC simulations of the gauge-glass XY model, for examples
$1/\nu=0.39(3)$ and $1/\nu=0.36(3)$ obtained in \cite{KY-02} and
\cite{NW-04} respectively, and by numerical calculations of the stiffness
exponent at $T=0$, for example $1/\nu=0.36(1)$ and $1/\nu\approx 0.45$
obtained in \cite{KY-02} and \cite{TT-05}.  Our result for $\eta_o$ is
consistent with a general argument which predicts $\eta_o = 0$.
\end{itemize}
 
\appendix

\section{Details on the Monte Carlo simulation} \label{AppMC}

In the simulation we use both Metropolis and microcanonical local updates. 
The latter  do not change the energy of the configuration and are defined
as follows. Consider a site $i$; the corresponding field is 
$\psi_i$. The terms of the Hamiltonian that depend on 
$\psi_i$ can be written as 
\begin{equation}
{\cal H}_i = {\rm Re}\, (\overline{\psi}_i z), \qquad
\qquad z \equiv \sum_j U_{ij} \psi_j ,
\end{equation}
where the sum is over all nearest neighbors $j$ of site $i$. 
Then, define 
\begin{equation}
\psi'_i = 2 {z\over |z|^2}{\rm Re}\, (\overline{\psi}_i z)  - \psi_i
\end{equation}
One can verify that $|\psi'_i| = 1$ and that 
\begin{equation}
{\rm Re}\, (\overline{\psi}_i z) = {\rm Re}\, (\overline{\psi}_i' z).
\end{equation}
Thus, the update $\psi_i\to\psi_i'$ does not change the energy and can 
therefore be always accepted. This update does not suffer the limitations
of the Metropolis update: $\psi_i$ and $\psi_i'$ are not close to each other.

In our simulation a MC step consists of 5 microcanonical sweeps over all the 
lattice followed by one Metropolis sweep. For each disorder sample we 
typically perform $O(10^5)$ MC steps. In some simulations of the CRPXY
model we also use the parallel tempering method ~\cite{raex,par-temp}.
It allows us to obtain results for small values of $T$, in particular
below the Nishimori line $T = \sigma$.
In the parallel-tempering
simulations we consider $N_T$ systems at the same value of
$\sigma$ and at $N_T$ different inverse temperatures $\beta_{\rm min} \equiv
\beta_1$, \ldots, $\beta_{\rm max}$, where $\beta_{\rm max}$ corresponds to
the minimum value of the temperature we are interested in.  The value
$\beta_{\rm min}$ is chosen so that thermalization at $\beta=\beta_{\rm min}$
is sufficiently fast, while  
the intermediate values $\beta_i$ are chosen so that
the acceptance probability of the temperature exchange is at least
$5\%$.  Moreover,  we require that, for some $i$, $\beta_i = \sigma$. 
This allows us to collect data on the Nishimori line.
The exact results valid on it allow us to check the correctness of the 
MC code and perform a (weak) test of thermalization. Thermalization is 
checked by verifying that the averages of the observables are independent 
of the number of MC steps for each disorder realization.

The overlap correlations and the corresponding $\chi_o$ and $\xi_o$ are 
measured by performing two independent runs for each disorder sample.  
Finally, note that the determination of $g_{22}$ defined in (\ref{g22def}) 
requires the computation of the disorder average of products of 
thermal expectations.
This should be done with care in order to avoid any bias due to 
the finite length of the run for each disorder realization. 
We use the essentially unbiased estimators discussed in 
\cite{HPPV-07-bias,HPV-08}.

\section{The KT RG equations} \label{AppRGsigmazero}

In this Appendix we consider the RG flow for the sine Gordon (SG) model,
with the purpose of understanding its universal features. As 
a results we shall obtain the critical behavior of the correlation length
and of the magnetic susceptibility at the KT transition. 
This appendix generalizes the results presented in 
\cite{AGG-80,BH-00,Balog-01}. 
The SG model is parametrized by two couplings,
$\alpha$ and $\delta$---we use the notations of 
\cite{AGG-80,Balog-01}---whose $\beta$ functions are
\begin{eqnarray}
&&\beta_\alpha = 2\alpha\delta + {5\over 64} \alpha^3 + \ldots,\\
&&\beta_\delta = {1\over 32} \alpha^2 - {1\over 16} \alpha^2\delta + \ldots,
\end{eqnarray}
where the dots indicate higher-order terms. To all orders the 
$\beta$ functions have the generic form
\begin{eqnarray}
&&\beta_\alpha = 2\alpha\delta + 
   \sum_{n+m>2} b_{\alpha,nm} \alpha^n \delta^m, \\
&& \beta_\delta = {1\over 32} \alpha^2 + 
   \sum_{n+m>2} b_{\delta,nm} \alpha^n \delta^m.
\end{eqnarray}
In the SG model the sign of $\alpha$ is irrelevant, which implies the symmetry
relations
\begin{equation}
\beta_\alpha(\alpha,\delta) = - \beta_\alpha(-\alpha,\delta), \qquad\qquad
\beta_\delta(\alpha,\delta) =  \beta_\delta(-\alpha,\delta). \qquad\qquad
\end{equation}
As a consequence, $b_{\alpha,nm} = 0$ if $n$ is even and $b_{\delta,nm} = 0$
if $n$ is odd. Moreover, for $\alpha = 0$ the theory is free and $\delta$ does
not flow. Hence
\begin{equation}
\beta_\delta(\alpha=0,\delta)  = 0,
\end{equation}
which implies $b_{\delta,nm} = 0$ if $n=0$.

Let us now consider a general nonlinear analytic redefinition of the couplings
\begin{eqnarray}
\alpha &=& a_{\alpha,10} u + 
   \sum_{n+m\ge 2} a_{\alpha,nm} u^n v^m, \\
\delta &=& a_{\delta,01} v + 
   \sum_{n+m\ge 2} a_{\delta,nm} u^n v^m.
\end{eqnarray}
We have verified up to the $7^{\rm th}$ order that with a proper choice of the
coefficients $a_{\alpha,nm}$ and $a_{\delta,nm}$ one can rewrite the $\beta$
functions in the form
\begin{eqnarray}
\beta_u(u,v) &=& - u v, 
\label{betau}\\
\beta_v(u,v) &=& - u^2 (1 + b_1 v + b_3 v^3 + b_5 v^5 + \ldots).
\label{betav-1} 
\end{eqnarray}
The couplings $u$ and $v$ are not uniquely defined and indeed there is a
family of transformations that do not change the $\beta$ functions
(\ref{betau}) and (\ref{betav-1}).  Extending the previous results to all
orders, in the following we assume that we can choose $u$ and $v$ in such a
way that $\beta_u(u,v)$ is given by (\ref{betau}) and $\beta_v(u,v)$ has
the form
\begin{equation}
\beta_v(u,v) = - u^2 [1 + v f(v^2)],
\label{betav} 
\end{equation}
where $f(v^2)$ is an analytic function in the region $v<v_0$, where $v_0$ is
the starting point of the RG flow, and satisfies $1 + v f(v^2)>0$ in this
domain (if this were not true, we would have another nontrivial fixed point).
This parametrization is unique (universal) in the sense that there is no
analytic redefinition of the couplings which allows one to write the $\beta$
functions in the form (\ref{betau}), (\ref{betav}) with a different function
$f(v^2)$, i.e. with different coefficients $b_{2n+1}$. The perturbative
calculations of \cite{AGG-80} allow us to determine $b_1$:
\begin{equation}
b_1 = -{3\over2}.
\label{b1calc}
\end{equation}
The analysis of the flow in the general case is analogous to that presented in
\cite{AGG-80,Balog-01}. First, we define the RG invariant function
\begin{eqnarray}
&&Q(u,v) = u^2 - F(v),
\label{qdef} \\\
&&F(v) = 2 \int_0^v {w dw\over 1 + w f(w^2)} = v^2 + v^3 + {9\over 8} v^4
+ O(v^5),
\nonumber
\end{eqnarray}
which satisfies 
\begin{equation}
{dQ\over dl} = {\partial Q\over \partial u} \beta_u(u,v) + 
   {\partial Q\over \partial v} \beta_v(u,v)  = 0,
\end{equation}
where $l$ is the flow parameter. The RG flow follows the lines $Q =\,
$constant.  It is thus natural to parametrize the RG flow in terms of $Q$ and
$v(l)$.  Since
\begin{equation}
{dv\over dl} = \beta_v(u,v) = - [Q + F(v)][1 + v f(v^2)],
\label{betavqf}
\end{equation}
we obtain 
\begin{equation}
l = - \int_{v_0}^v {dw\over [Q + F(w)][1 + w f(w^2)]},
\end{equation}
where $v(l=0) = v_0$.

Let us now apply these results to the XY model. Repeating the discussion of
\cite{Kosterlitz-74,JKKN-78} the XY model can be mapped onto a line in
the $(u,v)$ plane with $v > 0$.  The KT transition is the intersection of this
line with the line $Q=0$ and the high-temperature phase corresponds to $Q >
0$.  Thus, $Q$ plays the role of thermal nonlinear scaling field, i.e.
\begin{equation}
Q = q_1 \tau + q_2 \tau^2 + \ldots
\end{equation}
where $\tau = (T-T_{XY})/T_{XY}$. 

To derive the expected critical behavior we consider the singular part of the
free energy in a box of size $L$. It satisfies the scaling equation
\cite{Wegner-76}
\begin{equation}
{\cal F}_{\rm sing} (\tau,L) =     e^{-2l} f(Q,v(l),e^{-l}L)~,
\end{equation}
where we have parametrized the flow in terms of $Q$ and $v(l)$ and we have
neglected all irrelevant operators.  If $Q > 0$, as discussed in
\cite{AGG-80}, $v(l)$ decreases continuously and $v(l)\to -\infty$ as
$l\to \infty$.  Since $v_0$,
the starting point of the flow, is positive, we can fix $l$ be requiring
\begin{equation}
   v(l) = -1,
\end{equation}
so that 
\begin{equation}
l = \int^{v_0}_{-1} {dw\over [Q + F(w)][1 + w f(w^2)]} = I(Q,v_0).
\end{equation}
It follows
\begin{equation}
{\cal F}_{\rm sing}(\tau,L) = 
    e^{-2I(Q,v_0)} f(Q,-1,e^{-I(Q,v_0)}L)~,
\end{equation}
which gives the scaling behavior of the free energy (using $Q\sim
\tau$). In the scaling limit the finite-size dependence can be parametrized in
terms of $\xi/L$, where $\xi$ is the correlation length. This allows us to
identify
\begin{equation}
\xi(\tau) = \xi_0 e^{I(Q,v_0)}~,
\end{equation}
where $\xi_0$ is a constant.
The behavior of $\xi(\tau)$ for $\tau \to 0$ is obtained by expanding
$I(Q,v_0)$ for $Q\to 0$. The generic behavior is
\begin{equation}
I(Q,v_0) = {1\over \sqrt{Q}} \sum_n I_n Q^n + 
            \sum_n I_{{\rm an},n}(v_0) Q^n .
\end{equation}
The nonanalytic terms in the expansion depend only of the coefficients
$b_{2n+1}$ which appear in (\ref{betav-1}).  The first two coefficients
are
\begin{eqnarray}
I_0 &=& \pi, \nonumber \\
I_1 &=& {\pi b_1\over 4} = {9 \pi\over 16} .
\end{eqnarray}
Correspondingly, we obtain
\begin{equation}
\xi(\tau) = X \exp (\pi/\sqrt{Q}) [1 + I_1 \sqrt{Q} + O(Q)].
\label{xiKT-Q}
\end{equation}
Expanding $Q$ in powers of $\tau$ we obtain the celebrated KT expression for
the correlation length.

Let us now consider the behavior of the susceptibility.  Perturbation theory
gives for the scaling dimension of the spin correlation function \cite{AGG-80}
\begin{eqnarray}
&&\gamma = -{1\over 4} + {1\over 4}\delta - {1\over 4} \delta^2 +
h_1\alpha^2+ \ldots , 
\end{eqnarray}
where $h_1$ is an unknown coefficient. If we perform the redefinitions
$(\alpha,\delta) \to (u,v)$ considered before, we can rewrite $\gamma$ 
as\footnote{The possibility of cancelling the term of order $\alpha^2$ is
  related to the existence of a family of transformations 
  transformations, given at second order by $u' = u + A u v $, $v' =
  v + A u^2$ with arbitrary $A$, which leave invariant the 
  $\beta$-functions
  (\ref{betau}) and (\ref{betav-1}).  By properly choosing $A$ one can
  eliminate the $\alpha^2$ term in $\gamma(u',v')$.  }
\begin{eqnarray}
&&\gamma = -{1\over 4} - {1\over 8}v - {1\over 16} v^2 + \ldots
\label{gammapert}
\end{eqnarray}
without the $\alpha^2$ term.  In the infinite-volume limit the susceptibility
satisfies the scaling law
\begin{equation}
\chi\xi^{-7/4}  = A
    \exp\left[\int_{v_0}^{v(l)} {\gamma(w) + 1/4\over 
\beta_v} dw \right]
    G_\chi[Q,v(l)],
\label{chisuxi7/4-RG}
\end{equation}
the integral is computed at fixed $Q$ with $\beta_v$ given by
(\ref{betavqf}), and $G_\chi$ is an analytical function.  
Setting $v(l)=-1$ and expanding the integral in powers
of $Q$, we obtain an expansion of the form
\begin{equation}
\chi\xi^{-7/4}  = A(1 + c_1 \sqrt{Q} + c_2 Q + \ldots).
\end{equation}
The coefficient $c_1$ can be computed exactly using the perturbative results
(\ref{betav-1}), (\ref{b1calc}), and (\ref{gammapert}), obtaining
\begin{equation}
   c_1 = {\pi\over 16}.
\end{equation}
Using (\ref{xiKT-Q}) we can write
\begin{equation}
\sqrt{Q} = {\pi\over\ln \xi/X} + O(\ln^{-3}\xi)
\end{equation}
and obtain 
\begin{equation}
\chi\xi^{-7/4}  = A_\chi
   \left[1 + {\pi^2\over 16 \ln(\xi/X)} + O(1/\ln^{2}\xi) \right].
\label{chi-expRG}
\end{equation}
Note that the leading logarithmic scaling correction has a universal
coefficient.  We should note that in \cite{Balog-01} it was incorrectly
claimed that $c_1 = 0$ and, as a consequence, that the leading scaling
corrections in (\ref{chi-expRG}) are proportional to $1/(\ln\xi)^2$.  We
numerically checked (\ref{chi-expRG}) by fitting the infinite-volume
numerical data of \cite{BNNPSW-01} (more precisely their data for
$\beta\ge 0.92$, corresponding to $10\lesssim \xi\lesssim 420$) to
\begin{eqnarray}
\ln (\chi \xi^{-7/4}) = a + {b\over \ln(\xi/X)} ,\label{fita}
\end{eqnarray}
obtaining $a=0.804(2)$ and $b=0.627(9)$ (with $\chi^2/{\rm DOF}\approx 0.7$),
which is perfectly consistent with the value of $b$ obtained in perturbation
theory, i.e.  $b=\pi^2/16\approx 0.617$ (fixing $b=\pi^2/16$, we obtain
$a=0.8058(1)$ with $\chi^2/{\rm DOF}\approx 0.7$, while a fit to $a +
{b/\ln(\xi/X)} + {c/\ln^2(\xi/X)}$ gives $a=0.8046(9)$, $c=0.029(22)$ with
$\chi^2/{\rm DOF}\approx 0.6$, which confirms that the next-to-leading
correction is very small in (\ref{fita})).

The result (\ref{chi-expRG}) is general.  If ${\cal O}$ is a generic 
long-distance quantity
which behaves as $\xi_o^{x}$ in the critical limit, we expect ${\cal
  O}/\xi_o^{x}$ to behave as $\chi/\xi^{7/4}$, i.e. to satisfy a relation
analogous to (\ref{chisuxi7/4-RG}). It is only needed to replace
$\gamma(u,v) + 1/4$ with the appropriate subtracted scaling dimension.  
Thus, ${\cal O}/\xi_o^{x}$ also has an
expansion of the form (\ref{chi-expRG}), i.e.
\begin{equation}
{\cal O} = \xi_o^{x} \left[1 + {c_{\cal O}\over \ln \xi/X} + 
     O(\ln^{-2}\xi) \right],
\end{equation}
where $c_{\cal O}$ is universal and can be computed by using 
the perturbative expression of the scaling dimension of ${\cal O}$. 
More precisely, if the scaling dimension $\gamma_{\cal O}(u,v)$ has the 
perturbative expansion
\begin{equation}
\gamma_{\cal O}(u,v) = g_{00} + g_{01} v + g_{02} v^2 + g_{20} u^2 + \ldots
\end{equation}
we obtain
\begin{equation}
c_{\cal O} = - \pi g_{02}.
\end{equation}
Corrections proportional to $1/\ln \xi/X$ should instead be absent in RG
invariant quantities. Indeed, if $R$ is such a quantity, if we neglect the
scaling corrections, $R$ satisfies the scaling relation
\begin{equation}
R(\tau) = G_R[Q,v(l)],
\end{equation}
for any $l$. This implies that $R(\tau)$ is independent of $v(l)$,  
hence an analytic function of $Q$ and therefore of $\tau$. It follows
\begin{equation}
R(\tau) = R^* + {c_R\over\ln^2 \xi/X} + O(\ln^{-4}\xi),
\end{equation}
where the costant $c_R$ is expected to be universal.

\section{General behavior close to a critical point} \label{Appirrelevant}

Let us consider a multicritical point in a two-parameter space 
labelled by $T$ and $\sigma$ and let us assume that the correlation length
behaves as 
\begin{eqnarray}
\xi(T,\sigma) &\sim& [T-T_{c}(0)]^{-\nu_1} \qquad\qquad \sigma = 0, 
\label{xisigma0} \\
\xi(T,\sigma) &\sim& [T-T_{c}(\sigma)]^{-\nu_2} \qquad\qquad \sigma > 0, 
\label{xisigmaneq0} 
\end{eqnarray}
where $T_{c}(\sigma)$ is the $\sigma$-dependent critical point and 
$\nu_1 \not=\nu_2$. According to the RG, close to the 
multicritical point $\xi(T,\sigma)$ behaves as 
\begin{equation}
\xi(T,\sigma) = u_t(T,\sigma)^{-\nu_{m}} 
    F[u_\sigma(T,\sigma) u_t(T,\sigma)^{-\phi}],
\label{MCP}
\end{equation}
where $u_\sigma(T,\sigma)$ and $u_t(T,\sigma)$ are the scaling fields
and $\phi$ and $\nu_{m}$ two critical exponents.
Since one of the two scaling fields must vanish along the 
transition line, we define $u_t(T,\sigma)$ as the scaling 
field which has this property. Therefore, we define
\begin{eqnarray}
u_t(T,\sigma) = {T - T_c(\sigma) \over T_c(0)}~.
\end{eqnarray}
For $\sigma\to0$ and $T\to T_c(0)$, it behaves as 
\begin{equation}
u_t(T,\sigma) = \tau + c_\sigma \sigma + \ldots \qquad 
   \tau \equiv {T - T_c(0) \over T_c(0)}.
\end{equation}
We assume that $c_\sigma\not=0$, i.e. that the transition line is not
perpendicular to the line $\sigma =0$, as it occurs in the RPXY
model. Finally, we note that $u_\sigma(T,\sigma)$ does not vanish on the
transition line, unless $\sigma = 0$.

Now consider $T\to T_c(\sigma)$ at fixed novanishing $\sigma$.  Since
$u_\sigma(T,\sigma)\not=0$ we obtain (\ref{xisigmaneq0}) only if
\begin{equation}
F(x) \sim x^\lambda\qquad \qquad \lambda = {\nu_2 - \nu_{m}\over \phi}
\label{scalF}
\end{equation}
for $x \to \infty$. 
To go further let us distinguish two cases: (i) $u_\sigma(T,\sigma)$
vanishes identically for $\sigma=0$, i.e. $u_\sigma(T,0)=0$ for any $T$;
(ii) $u_\sigma(T,0)$ is different from zero unless $T=T_c(\sigma=0)$.

In case (i) (\ref{xisigma0}) requires
\begin{equation}
 F(0) \not=0, \qquad\qquad \nu_{m} = \nu_1.
\end{equation}
Assuming $u_\sigma(T=T_c(0),\sigma) = d_\sigma\sigma$ for $\sigma\to 0$ 
we obtain 
\begin{equation}
  \xi(T=T_c(0),\sigma) = (c_\sigma \sigma)^{-\nu_1} 
    F(d_\sigma c_\sigma^{-\phi} \sigma^{1-\phi}).
\end{equation}
The observed behavior depends on the value of $\phi$. For $\phi< 1$,
since $F(0) \not=0$ we obtain 
\begin{equation}
\xi(T=T_c(0),\sigma) = (c_\sigma \sigma)^{-\nu_1} (a + b \sigma^{1-\phi} + 
   \ldots )
\end{equation}
The corrections are correct provided that $F(x)$ is analytic for $x=0$.  If
$\phi > 1$, using (\ref{scalF}) we obtain the behavior
\begin{equation}
\xi(T=T_c(0),\sigma) \sim \sigma^{-\overline{\nu}} 
   \qquad \overline{\nu} = \nu_1 - (1-\phi)\lambda = 
    {\nu_2(\phi-1) + \nu_1\over \phi}~.
\end{equation}

In case (ii), 
if $u_\sigma(T,\sigma=0)=d_T \tau + O(\tau^2)$ we obtain for $\sigma = 0$
\begin{equation}
\xi(T,0) = \tau^{-\nu_{m}} F(d_T \tau^{1-\phi}),
\end{equation}
which shows that 
\begin{equation}
F(d_T \tau^{1-\phi}) \sim \tau^{\nu_m - \nu_1}
\label{scal-F-App}
\end{equation}
in the limit $\tau\to 0$.
Let us now consider the behavior for $T = T_c(0)$ as a function of $\sigma$.
For $\sigma\to 0$ we have
\begin{equation}
\xi(T=T_c(0),\sigma) = c_\sigma^{-\nu_m} \sigma^{-\nu_m}
    F(d_\sigma c_\sigma^{-\phi}\sigma^{1-\phi}) \sim \sigma^{-\nu_1},
\end{equation}
where we have used relation (\ref{scal-F-App}). Thus, in case (ii) we have
$\xi(T=T_c(0),\sigma)\sim \sigma^{-\nu_1}$ for any value of $\phi$.

Let us now show that the case relevant for the RPXY model is 
case (i). Indeed, case (ii) can only occur if the two relevant 
operators which occur at the
multicritical point are both present in the model at $\sigma = 0$. This does
certainly not occur in our case in which $\sigma$ is associated with
randomness. Therefore, our result that in the RPXY model
$\xi(T=T_c(0),\sigma)$ behaves as $\sigma^{-\nu_1}$ implies that $\phi <
1$, i.e. that the
RG dimension of the new operator that arises in the theory with $\sigma\not=0$
is less relevant than the thermal operator present at $\sigma = 0$.
This is also the case of three-dimensional randomly dilute Ising
systems or $\pm J$ Ising models at their ferromagnetic transitions at small
disorder. Indeed, the crossover from the pure critical behavior to that of the
randomly-dilute Ising universality class is described by the crossover
exponent $\phi=\alpha_{\rm Is}=0.1096(5)$~\cite{PV-r,CPRV-02}, see also the
discussion reported in \cite{HPPV-07}.

Similar considerations apply to other quantities. For instance, consider 
a RG invariant quantity $R$. It behaves as 
\begin{equation}
R(T,\sigma) = r[u_\sigma(T,\sigma) u_t(T,\sigma)^{-\phi}].
\end{equation}
If $\phi<1$, $R(T,\sigma)$ approaches the same value $R^*$ along the 
lines $\sigma = 0$  and $T=T_c(0)$. Morover, in the second case 
we expect corrections of the form
\begin{equation}
R(T_c(0),\sigma) =  R^* + a \sigma^{1-\phi} + \ldots= 
       R^* + a' \xi^{(\phi-1)/\nu_1} + \ldots
\end{equation}

\section{RG equations in the presence of randomness}
\label{AppRGdisordine}

The RG equations in the small disorder regime and close to the 
paramagnetic-QLRO transition line have been derived in 
\cite{RSN-83,NSKL-95,Tang-96,Scheidl-97,CL-98}:
\begin{eqnarray}
&&{d T \over dl} = - 4 \pi^3 Y^2, \nonumber \\
&& {d\sigma\over dl} = 0,\nonumber \\
&& {dY\over dl} = (2 - \pi \beta + \pi \sigma \beta^2) Y,
\nonumber
\end{eqnarray} 
where $Y$ is the vorticity and only terms up to $O(Y^2)$ are kept. Let us now
redefine the couplings as follows:
\begin{eqnarray}
&& T^{-1} = {1\over \pi} (2 + v + \sigma), \nonumber \\
&& Y  = {u\over 4\pi}.
\end{eqnarray}
For $u,v\to 0$ the RG equations become
\begin{eqnarray}
&& {du\over dl} = - uv, \nonumber \\
&& {dv\over dl} = - u^2, \nonumber \\
&& {d\sigma\over dl} = 0.
\label{RGeq-disorder}
\end{eqnarray}
We have thus reobtained the RG equations for the XY model. 
This implies that, in the region of couplings in which 
(\ref{RGeq-disorder}) hold, the RG behavior is analogous to 
that close to the KT fixed point, apart from an analytic 
redefinition of the scaling fields.

\section{Magnetic correlations in the gauge-glass model} 
\label{App-magn}

For the gauge-glass model ($\sigma=\infty$) we can derive some identities
which relate magnetic and overlap quantities. The basic observation is that 
for $\sigma = +\infty$ the distribution function of the 
$A_{xy}$ variables is gauge-invariant. Hence we have
\begin{eqnarray}
&& [\langle {\psi}_{x_1}^* \ldots {\psi}_{x_n}^*
   \psi_{y_1} \ldots \psi_{y_n} \rangle ] = 
 V^*_{x_1} \ldots V^*_{x_n} V_{y_1}  \ldots V_{y_n}
 [\langle {\psi}_{x_1}^* \ldots {\psi}_{x_n}^*
   \psi_{y_1} \ldots \psi_{y_n} \rangle ],
\end{eqnarray}
for any set of phases $V_x$.
It implies that magnetic correlations vanish
unless each $x_i$ is equal to some $y_j$. 
Analogously we have
\begin{eqnarray}
&& [\langle {\psi}_{x_1}^* \ldots {\psi}_{x_n}^*
   \psi_{y_1} \ldots \psi_{y_n} \rangle 
   \langle {\psi}_{z_1}^* \ldots {\psi}_{z_n}^*
   \psi_{t_1} \ldots \psi_{t_n} \rangle
] =  \\
&&  = 
 V^*_{x_1} \ldots V^*_{x_n} V_{y_1}  \ldots V_{y_n}
 V^*_{z_1} \ldots V^*_{z_n} V_{t_1}  \ldots V_{t_n} 
   [\langle {\psi}_{x_1}^* \ldots {\psi}_{x_n}^*
   \psi_{y_1} \ldots \psi_{y_n} \rangle
   \langle {\psi}_{z_1}^* \ldots {\psi}_{z_n}^*
   \psi_{t_1} \ldots \psi_{t_n} \rangle].
\nonumber 
\end{eqnarray}
These relations allow us to write 
\begin{eqnarray}
&& [\langle {\psi}_{x}^*  \psi_{y} \rangle ] = 
   \delta_{xy}, \\
&& [\langle {\psi}_{x_1}^* {\psi}_{x_2}^* \psi_{y_1} \psi_{y_2}\rangle ] = 
  \delta_{x_1y_1} \delta_{x_2y_2} + \delta_{x_1y_2} \delta_{x_2y_1} - 
  \delta_{x_1y_1} \delta_{x_1x_2} \delta_{x_1y_2},  \\
&& [\langle {\psi}_{x_1}^* \psi_{y_1} \rangle \langle 
         {\psi}_{x_2}^* \psi_{y_2}\rangle ] = 
   \delta_{x_1y_1} \delta_{x_2y_2} +
   \delta_{x_1y_2} \delta_{x_2y_1} 
    [|\langle {\psi}_{x_1}^* \psi_{y_1} \rangle|^2] - 
  \delta_{x_1y_1} \delta_{x_1x_2} \delta_{x_1y_2}.
\end{eqnarray}
It follows
\begin{eqnarray}
&& [\langle | \mu |^2\rangle ] = V, \nonumber \\
&& [\langle | \mu |^4\rangle ] = 2 V^2 - V, \nonumber \\
&& [\langle | \mu |^2\rangle^2 ] = V^2 + V^2 \chi_o - V,
\end{eqnarray}
which imply
\begin{eqnarray}
&& \chi = 1,  \nonumber \\
&& \chi_4 = 1 - 2 \chi_o, \nonumber \\
&& \chi_{22} = \chi_o  - 1.
\label{refchi}
\end{eqnarray}
Moreover, it is easy to show that $\xi = 0$.
Relations (\ref{refchi})
show that $\chi_4$ and $\chi_{22}$ both diverge as $\chi_o$.
In the critical limit we have $\chi_o\sim \xi_o^2$ because
$\eta_o = 0$. Therefore we can write
\begin{equation}
\chi_4 \approx - 2 a \xi_o^2, \qquad\qquad 
   \chi_{22} \approx a \xi_o^2, 
\label{relchi}
\end{equation}
for $\xi_o\to \infty$, where $a$ is constant.

We shall now assume that these results are valid for the whole 
universality class: for any $\sigma > \sigma_D$, 
relations (\ref{relchi}) always hold with a constant $a$ which
in general depends on $\sigma$. 
We can reexpress these results in terms of the quartic couplings.
If we use (\ref{relchi}) we have 
\begin{eqnarray}
g_4 &=& {3 a \xi_o^2\over \chi^2 \xi^2}, \\
g_{22} &=&  -{  a \xi_o^2\over \chi^2 \xi^2}, 
%\\g_c &=& {3\over 2 \xi^2}.
\label{as-gc}
\end{eqnarray}
Since the magnetic susceptibility $\chi$ and correlation length $\xi$ are
finite and nonzero (except for $\sigma=\infty$, where anyhow the quartic
couplings are not well-defined since $\xi=0$ for any $L$), we expect that
$g_4$ and $g_{22}$ diverge as $\xi_o^2$ in the critical limit.  As for
$g_c=g_4+3g_{22}$, (\ref{as-gc}) shows that the leading $\xi^2_o$ term
cancels. Since in the calculation we have neglected the scaling corrections to
(\ref{relchi}), this does not necessarily imply that $g_c$ remains finite
in the critical limit, but only that $g_c \xi_o^{-2} \to 0$ as
$\xi_o\to\infty$. The exact behavior depends on the neglected scaling
corrections.  These predictions are confirmed by our numerical results, see
Sec.~\ref{g4g22magn}.  It is worth mentioning that this behavior is analogous
to that observed in the 2D Ising spin glass model, where $\chi_4$
behaves as $\chi_o$ and
thus diverges approaching the glassy transition; see, e.g.,
\cite{BY-86} and references therein.


\section*{References}
 \begin{thebibliography}{99}

\bibitem{GK-86}
Granato E and Kosterlitz J M 1986 
Quenched disorder in Josephson-junction arrays in a transverse magnetic field
{\em Phys. Rev.} B {\bf 33} 6533 

\bibitem{GK-89}
Granato E and Kosterlitz J M 1989
Disorder in Josephson-junction arrays in a magnetic field
{\em Phys. Rev. Lett.} {\bf 62} 823

\bibitem{RSN-83}
Rubinstein M, Shrainam B and Nelson D R 1983
Two-dimensional XY magnets with random Dzyaloshinskii-Moriya interactions
{\em Phys. Rev.} B {\bf 27} 1800 

\bibitem{CF-94}
Cha M-C and Fertig H A 1994
Orientational order and depinning of the disordered electron solid
{\em Phys. Rev. Lett.} {\bf 73} 870 
[arXiv:cond-mat/9402021] \par
Cha M-C and Fertig H A 1994
Topological defects, orientational order, 
and depinning of the electron solid in a random potential
{\em Phys. Rev.} {\bf B} 50 14368 
[arXiv:cond-mat/9409001]

\bibitem{FTY-91}
Fisher M P A, Tokuyasu T A and Young A P 1991
Vortex variable-range-hopping resistivity in superconducting films
{\em Phys. Rev. Lett.} {\bf 66} 2931

\bibitem{Korshunov-06}
Korshunov S E 2006
Phase transitions in two-dimensional systems with continuous degeneracy
{\em Usp. Fiz. Nauk} {\bf 176} 233,
{\em Physics Uspekhi} {\bf 49} 225 (English translation)

\bibitem{KR-03}
Kawashima N and Rieger H 2004 
Recent Progress in Spin Glasses
in {\em Frustrated Spin Systems} ed H T  Diep
(World Scientific: Singapore) 
[arXiv:cond-mat/0312432]

\bibitem{ON-93}
Ozeki Y and Nishimori H 1993
Phase diagram of gauge glasses
{\em J. Phys.} A: {\em Math. Gen.} {\bf 26} 3399

\bibitem{Nishimori-02}
Nishimori H 2002 
Exact results on spin glass models 
{\em Physica} A {\bf 306} 68
[arXiv:cond-mat/0201056]

\bibitem{ES-85}
Ebner C and Stroud D 1985
Diamagnetic susceptibility of superconducting clusters: Spin-glass behavior
{\em Phys. Rev.} B {\bf 31}  165

\bibitem{FLTL-88}
Forrester M G, Lee Hu Jong, Tinkhams M and Lobb C J 1988
Positional disorder in Josephson-junction arrays: Experiments and simulations
{\em Phys. Rev.} B {\bf 37} 5966 

\bibitem{CD-88}
Chakrabarti A and Dasgupta C 1988
Phase transition in positionally disordered Josephson-junction arrays in a 
transverse magnetic field
{\em Phys. Rev.} B {\bf 37} 7557

\bibitem{FBL-90}
Forrester M G, Benz S P and  Lobb C J 1990
Monte Carlo simulations of Josephson-junction arrays with positional disorder
{\em Phys. Rev.} B {\bf 41} 8749 

\bibitem{HS-90}
Huse D A and Seung H S 1990
Possible vortex-glass transition in a model random superconductor
{\em Phys. Rev.} B {\bf 42} 1059 

\bibitem{RTYF-91}
Reger J D, Tokuyasu T A, Young A P and Fisher M P A 1991
Vortex-glass transition in three dimensions
{\em Phys. Rev.} B {\bf 44} 7147

\bibitem{Li-92}
Li Y-H 1992 
Voltage-current characteristics of the two-dimensional gauge glass model
{\em Phys. Rev. Lett.} {\bf 69} 1819

\bibitem{Gingras-92}
Gingras M J P 1992 
Numerical study of vortex-glass order in random-superconductor and 
related spin-glass models
{\em Phys. Rev.} B {\bf 45} 7547

\bibitem{DWKHG-92}
Dekker C, W\"oltgens P J M, Koch R H, Hussey B W and Gupta A 1992
Absence of a finite-temperature vortex-glass phase transition in 
two-dimensional YBa$_2$Cu$_3$O$_{7-\delta}$´ films
{\em Phys. Rev. Lett.} {\bf 69} 2717

\bibitem{RY-93}
Reger J D and Young A P 1993
Monte Carlo study of a vortex glass model
{\em J. Phys.} A: {\em Math. Gen.} {\bf 26} L1067
[arXiv:cond-mat/9311036]

\bibitem{Korshunov-93}
Korshunov S E 1993
Possible destruction of the ordered phase in Josephson-junction arrays 
with positional disorder
{\em  Phys. Rev.} B {\bf 48} 1124

\bibitem{NK-93}
Nishimori H and Kawamura H 1993
Gauge glass ordering in two dimensions
{\em J. Phys. Soc. Jpn.} {\bf 62} 3266

\bibitem{Nishimori-94}
Nishimori H 1994 
Gauge glass, spin glass and coding theory: Exact results
{\em Physica} A {\bf 205} 1

\bibitem{NSKL-95}
Nattermann T, Scheidl S, Korshunov S E and Li M S 1995
Absence of reentrance in the two-dimensional XY model with 
random phase shifts
{\em J. Physique I (France)} {\bf 5} 565
[arXiv:cond-mat/9501120]

\bibitem{CF-95}
Cha M-C and Fertig H A 1995
Disorder-induced phase transitions in two-dimensional crystals
{\em Phys. Rev. Lett.} {\bf 74} 4867

\bibitem{JKC-95}
Jeon G S, Kim S and Choi M Y 1995
Phase transition in the XY gauge glass
{\em Phys. Rev.} B {\bf 51} 16211

\bibitem{HWFGY-95}
Hyman R A, Wallin M, Fisher M P A, Girvin S M and Young A P 1995
Current-voltage characteristics of two-dimensional vortex-glass models
{\em Phys. Rev.} B {\bf 51} 15304
[arXiv:cond-mat/9409117]

\bibitem{KN-96}
Korshunov S E and Nattermann T 1996
Absence of reentrance in superconducting arrays with positional disorder
{\em Phys. Rev.} B {\bf 53} 2746

\bibitem{Tang-96}
Tang L-H 1996
Vortex statistics in a disordered two-dimensional XY model
{\em  Phys. Rev.} B {\bf 54} 3350
[arXiv:cond-mat/9602162]

\bibitem{BY-96}
Bokil H S and Young A P 1996
Study of chirality in the two-dimensional XY spin glass
{\em J. Phys.} A: {\em Math. Gen.} {\bf 29} L89
[arXiv:cond-mat/9512042]

\bibitem{MG-97}
Maucourt J and Grempel D R 1997
Phase transitions in the two-dimensional XY model with random phases: 
A Monte Carlo study.
{\em Phys. Rev.} B {\bf 56} 2572
[arXiv:cond-mat/9703109]

\bibitem{Scheidl-97}
Scheidl S 1997 
Glassy vortex state in a two-dimensional disordered XY model
{\em Phys. Rev.} B {\bf 55} 457
[arXiv:cond-mat/9601131]

\bibitem{KS-97}
Kosterlitz J M and Simkin M V 1997
Numerical study of a superconducting glass model
{\em Phys. Rev. Lett.} {\bf 79} 1098
[arXiv:cond-mat/9702166]

\bibitem{KCRS-97}
Kim B J, Choi M Y, Ryu S and Stroud D 1997
Anomalous relaxation in the XY gauge glass
{\em Phys. Rev.} B {\bf 56} 6007
[arXiv:cond-mat/9707140]

\bibitem{MG-98}
Maucourt J and Grempel D R 1998
Scaling of domain-wall energies in the three-dimensional gauge glass model
{\em Phys. Rev.} B {\bf 58} 2654 

\bibitem{CL-98}
Carpentier D and Le Doussal P 1998
Disordered XY models and Coulomb gases: 
Renormalization via traveling waves
{\em Phys. Rev. Lett.} {\bf 81} 2558
[arXiv:cond-mat/9802083]

\bibitem{Granato-98}
Granato E 1998 
Current-voltage scaling of chiral and gauge-glass models of two-dimensional 
superconductors
{\em Phys. Rev.} B {\bf 58} 11161
[arXiv:cond-mat/9808331]

\bibitem{SYMOUK-98}
Sawa A, Yamasaki H, Mawatari Y, Obara H, Umeda M and Kosaka S  1998
Thickness dependence of the vortex-glass transition and critical scaling of 
current-voltage characteristics in YBa$_2$Cu$_3$O$_{7-\delta}$´  thin films
{\em Phys. Rev.} B {\bf 58} 2868

\bibitem{KA-99}
Kosterlitz J M and Akino N 1999
Numerical study of spin and chiral order in a two-dimensional XY spin glass
{\em Phys. Rev. Lett.} {\bf 82} 4094
[arXiv:cond-mat/9806339]

\bibitem{MW-99}
Mudry C and Wen X-G 1999
Does quasi-long-range order in the two-dimensional XY model really survive 
weak random phase fluctuations? 
{\em Nucl. Phys.} B {\bf 549} 613
[arXiv:cond-mat/9712146]

\bibitem{CP-99}
Choi M Y and Park S Y 1999
Phase transition in the two-dimensional gauge glass
{\em Phys. Rev.} B {\bf 60} 4070
[arXiv:cond-mat/9906326]

\bibitem{Kim-00}
Kim B J 2000
Finite-temperature resistive transition in the two-dimensional XY gauge glass 
model
{\em Phys. Rev.} B {\bf 62} 644
[arXiv:cond-mat/0004069]

\bibitem{CL-00}
Carpentier D and Le Doussal P 2000
Topological transitions and freezing in XY models and Coulomb gases with 
quenched disorder: renormalization via traveling waves 
{\em Nucl. Phys.} B {\bf 588} 565
[arXiv:cond-mat/9908335]

\bibitem{AK-02}
Akino N and Kosterlitz J M 2002
Domain wall renormalization group study of the XY model with quenched 
random phase shifts
{\em Phys. Rev.} B {\bf 66} 054536
[arXiv:cond-mat/0203299]

\bibitem{HO-02}
Holme P and Olsson P 2002 
A zero-temperature study of vortex mobility in two-dimensional vortex glass 
models 
{\em Europhys. Lett.} {\bf 60} 439
[arXiv:cond-mat/0111555]

\bibitem{KY-02}
Katzgraber H G and Young A P  2002
Numerical studies of the two- and three-dimensional gauge glass at 
low temperature
{\em Phys. Rev.} B {\bf 66} 224507
[arXiv:cond-mat/0205206]

\bibitem{Katzgraber-03}
Katzgraber H G 2003
On the existence of a finite-temperature transition in the two-dimensional 
gauge glass
{\em Phys. Rev.} B {\bf 67} 180402(R)
[arXiv:cond-mat/0305393]
\par
Katzgraber H G 2003
Numerical studies of the two- and three-dimensional gauge glass at 
low temperature
{\em J. Applied Phys.} {\bf 93} 7661
[arXiv:cond-mat/0304540]

\bibitem{HKM-03}
Holme P, Kim B J and Minnhagen P 2003
Phase transitions in the two-dimensional random gauge XY model
{\em Phys. Rev.} B {\bf 67} 104510
[arXiv:cond-mat/0301279]

\bibitem{CTH-03}
Chen Q-H, Tanaka A and Hu X 2003
Evidence for finite-temperature glass transition in two dimensions
{\em Physica} B {\bf 329} 1413

\bibitem{NW-04}
Nikolaou M and Wallin M 2004
Zero-temperature glass transition in the two-dimensional gauge glass model
{\em Phys. Rev.} B {\bf 69} 184512
[arXiv:cond-mat/0312066]

\bibitem{KC-05}
Katzgraber H G and Campbell I A 2005
Dynamical scaling in Ising and vector spin glasses
{\em Phys. Rev.} B {\bf 72} 014462
[arXiv:cond-mat/0504082]

\bibitem{TT-05}
Tang L-H and Tong P 2005
Zero-temperature criticality in the two-dimensional gauge glass model
{\em Phys. Rev. Lett.} {\bf 94} 207204
[arXiv:cond-mat/0412415]

\bibitem{UKMCL-06}
Um J, Kim B J, Minnaghen P, Choi M Y and Lee S-I 2006
Dynamic critical behaviors in two-dimensional Josephson junction arrays with 
positional disorder
{\em Phys. Rev.} B {\bf 74} 094516
[arXiv:cond-mat/0608390]

\bibitem{YBC-06}
Yun Y J, Baek I C and Choi M Y 2006
Experimental study of positionally disordered Josephson junction arrays 
{\em Europhys. Lett.} {\bf 76} 271
[arXiv:cond-mat/0509151]

\bibitem{CLL-08}
Chen Q-H, Lv J-P and Liu H 2008
Dynamics of glass phases in the two-dimensional gauge glass model
{\em Phys. Rev.} B {\bf 78} 054519
[arXiv:0812.2822]

\bibitem{APV-09}
Alba V, Pelissetto A and Vicari E 2009  
Quasi-long-range order in the 2D XY model with random phase shifts
{\em J. Phys.: Math. Gen.} A {\bf 42} 295001
[arXiv:0901.4682]

\bibitem{KT-73} 
Kosterlitz J M and  Thouless D J 1973
Ordering, metastability and phase transitions in two-dimensional systems
{\em J.\ Phys.} C: {\em Solid State} {\bf 6}  1181

\bibitem{HP-97}
Hasenbusch M and  Pinn K 1997
Computing the roughening transition of Ising and 
solid-on-solid models by BCSOS model matching
{\em J. Phys.} A: {\em Math. Gen.} {\bf 30} 63 
[arXiv:cond-mat/9605019] \par
Hasenbusch M, Marcu M and Pinn K 1994
High-precision renormalization-group study of the roughening transition
{\em Physica} A {\bf 208} 124
[arXiv:hep-lat/9404016]

\bibitem{Nishimori-81}
Nishimori H 1981
Internal energy, specific heat and correlation function of the bond-random 
Ising model
{\em Prog. Theor. Phys.} {\bf 66} 1169

\bibitem{HPPV-08}
Hasenbusch M, Parisen Toldin F, Pelissetto A and Vicari E 2008 
Multicritical Nishimori point in the phase diagram of the  $\pm J$
Ising model on a square lattice
{\em Phys. Rev.} E {\bf 77} 051115
[arXiv:0803.0444]

\bibitem{PHP-06}
Picco M, Honecker A and Pujol P 2006
Strong disorder fixed points in the two-dimensional random-bond Ising model
{\em J. Stat. Mech.: Theory Exp.} P09006
[arXiv:cond-mat/0606312]

\bibitem{HPPV-08-2}
Hasenbusch M, Parisen Toldin F, Pelissetto A and Vicari E 2008
Universal dependence on disorder of two-dimensional randomly diluted and 
random-bond  $\pm J$ Ising models
{\em Phys. Rev.} E {\bf 78} 011110
[arXiv:0804.2788]

\bibitem{PPV-08}
Parisen Toldin F, Pelissetto A and Vicari E 2009
Strong-Disorder Paramagnetic-Ferromagnetic Fixed Point in the Square-Lattice 
$\pm J$ Ising Model 
{\em J. Stat. Phys.} {\bf 135} 1039
[arXiv:0811.2101]

\bibitem{PPV-10}
Parisen Toldin F, Pelissetto A and Vicari E 2010
in preparation.

\bibitem{AH-04}
Amoruso C and Hartmann A K 2004 
Domain-wall energies and magnetization of the two-dimensional random-bond 
Ising model
{\em Phys. Rev.} B {\bf 70} 134425 
[arXiv:cond-mat/0401464]

\bibitem{WHP-03}
Wang C, Harrington J and Preskill J 2003
Confinement-Higgs transition in a disordered gauge theory and the 
accuracy threshold for quantum memory 
{\em Ann. Phys. (NY)} {\bf 303} 31
[arXiv:quant-ph/0207088]

\bibitem{CPRV-96}
Campostrini M, Pelissetto A, Rossi P and Vicari E 1996
Strong-coupling analysis of two-dimensional O(N) models with $N\le2$
on square, triangular, and honeycomb lattices
{\em Phys. Rev.} B {\bf 54} 7301
[arXiv:hep-lat/9603002]

\bibitem{BNNPSW-01}
Balog J, Niedermaier M, Niedermayer F, 
Patrascioiu A, Seiler E and Weisz P 2001
Does the XY model have an integrable continuum limit? 
{\em Nucl. Phys.} B {\bf 618} 315
[arXiv:hep-lat/0106015]

\bibitem{PV-00}
Pelissetto A and Vicari E 2000
The effective potential of N-vector models: a field-theoretic study to 
$O(\epsilon^3)$
{\em Nucl. Phys.} B {\bf 575} 579 
[arXiv:cond-mat/9911452] \par
Pelissetto A and Vicari E 1998
Four-point renormalized coupling constant and Callan-Symanzik $\beta$-function 
in $O(N)$ models
{\em Nucl. Phys.} B {\bf 519} 123
[arXiv:cond-mat/9711078]

\bibitem{PV-r}
Pelissetto A and Vicari E 2002
Critical phenomena and renormalization-group theory
{\em Phys. Rep.} {\bf 368} 549
[arXiv:cond-mat/0012164]

\bibitem{raex} Geyer C J 1991
Markov chain Monte Carlo maximum likelihood
in {\em Computer Science and Statistics: Proc. of
    the 23rd Symposium on the Interface}, 
    ed E M Keramidas, p~156
  (Interface Foundation: Fairfax Station, VA, USA) \par
Hukushima K and Nemoto K 1996 
Exchange Monte Carlo method and application to spin glass simulations 
{\em J. Phys. Soc. Jpn.} {\bf 65} 1604

\bibitem{par-temp}
Earl D J and  Deem M W 2005
Parallel tempering: Theory, applications, and new perspectives 
{\em Phys. Chem. Chem. Phys.} {\bf 7} 3910
[arXiv:physics/0508111]

\bibitem{HPV-08}
Hasenbusch M, Pelissetto A and Vicari E 2008 
Critical behavior of three-dimensional Ising spin glass models
{\em Phys. Rev.} B {\bf 78} 214205 
[arXiv:0809.3329] \par
Hasenbusch M, Pelissetto A and Vicari E 2008 
The critical behavior of 3D Ising spin glass models: 
Universality and scaling corrections 
{\em J. Stat. Mech.: Theory Expt.} L02001
[arXiv:0710.1980]

\bibitem{HPPV-07-bias}
Hasenbusch M, Parisen Toldin F, Pelissetto A and Vicari E 2007
The universality class of 3D site-diluted and bond-diluted Ising systems
{\em J. Stat. Mech.: Theory Exp.} P02016
[arXiv:cond-mat/0611707]


\bibitem{AGG-80}
Amit D J, Goldschmidt Y Y and Grinstein G 1980
Renormalisation group analysis of the phase transition in the 2D Coulomb gas, 
Sine-Gordon theory and XY-model
{\em J. Phys.: Math. Gen.} A {\bf 13} 585

\bibitem{BH-00}
Balog J and Heged\~us A 2000 
Two-loop beta-functions of the sine-Gordon model
{\em J. Phys.: Math. Gen.} A {\bf 33} 6543
[arXiv:hep-th/0003258]

\bibitem{Balog-01}
Balog J 2001 
Kosterlitz-Thouless theory and lattice artifacts 
{\em J. Phys.: Math. Gen.} A {\bf 34} 5237
[arXiv:hep-lat/0011078]

\bibitem{Kosterlitz-74}
Kosterlitz J M 1974 
The critical properties of the two-dimensional $xy$ model
{\em J. Phys. C: Solid State Phys.} {\bf 7} 1046

\bibitem{JKKN-78}
Jos\'e J V, Kadanoff L P, Kirkpatrick S and Nelson D R 1978
Renormalization, vortices, and symmetry-breaking perturbations in 
the two-dimensional planar model
{\em Phys. Rev.} B {\bf 16} 1217

\bibitem{Wegner-76}
Wegner F J 1976
The Critical State, General Aspects
in Phase Transitions and Critical Phenomena,
Vol 6, ed C Domb and M Green, p 7
(Academic: New York) 

\bibitem{CPRV-02}
Campostrini M, Pelissetto A, Rossi P and Vicari E 2002
25th-order high-temperature expansion results for three-dimensional 
Ising-like systems on the simple-cubic lattice
{\em Phys. Rev.} E {\bf 65} 066127
[arXiv:cond-mat/0201180]

\bibitem{HPPV-07} 
Hasenbusch M, Parisen Toldin F, Pelissetto A  and Vicari E 2007
Magnetic-glassy multicritical behavior of the three-dimensional  
$\pm J$ Ising model
{\em Phys. Rev.} B {\bf 76} 184202
[arXiv:0707.2866]

\bibitem{BY-86}
Binder K and Young A P 1986
Spin glasses: Experimental facts, theoretical concepts, and open questions 
{\em Rev. Mod. Phys.} {\bf 58} 801
\end{thebibliography}
\end{document}